\documentclass[preprint]{aastex}
\usepackage{natbib}
\usepackage{color}
\usepackage{amsmath}
\usepackage{graphics}
\usepackage{tikz}
\usetikzlibrary{shapes,arrows}
\usetikzlibrary{shapes,snakes}
\begin{document}

\title{Visibility Estimation for the CHARA/JouFLU Exozodi Survey}

\author{Paul D. Nu\~nez\thanks{E-mail: paul.nunez@jpl.nasa.gov}, Theo ten Brummelaar$^2$, Bertrand Mennesson$^3$, Nicholas J. Scott$^2$ \\
$^{1}$NASA Postdoctoral Program Fellow, Jet Propulsion Laboratory, California Institute of Technology, 4800 Oak Grove Drive, Pasadena,\\ CA, 91109, USA.\\
$^{2}$The CHARA Array, Mount Wilson Observatory, Mount Wilson,\\ CA 91023, USA.
$^{3}$Jet Propulsion Laboratory, California Institute of Technology, 4800 Oak Grove Drive, Pasadena,\\ CA, 91109, USA.\\}

\begin{abstract}
  We discuss the estimation of the interferometric visibility (fringe contrast) for the exozodi survey conducted at the CHARA array with the JouFLU beam combiner. We investigate the use of the statistical median to estimate the \textit{uncalibrated} visibility from an ensemble of fringe exposures. Under a broad range of operating conditions, numerical simulations indicate that this estimator has a smaller bias  compared to other estimators. We also propose an improved method for \textit{calibrating} visibilities, which not only takes into account the time-interval between observations of calibrators and science targets, but also the uncertainties of the calibrators' raw visibilities. We test our methods with data corresponding to stars that do not display the exozodi phenomenon. The results of our tests show that the proposed method yields smaller biases and errors. The relative reduction in bias and error is generally modest, but can be as high as $\sim 20-40\%$ for the brightest stars of the CHARA data, and statistically significant at the $95\%$ confidence level (CL).  
\end{abstract}

\maketitle

\section{Introduction}

Two-beam optical interferometers have measured hundreds of angular diameters as small as a fraction of a milliarcsecond, with uncertainties of a few percent e.g. \citep{di_folco_2004, richichi, taby1, taby2, white_2013}, and have furthered our understanding of stellar structure and evolution. The improvements in precision of these instruments e.g. \citep{perrin_temp, fluor0, fluor2, merand_2006, pionier, nic} have also allowed detecting faint ($\sim 1\%$) near-infrared (NIR) exozodiacal light emitted from the vicinity ($0.1\,AU-10\,AU$) of stellar photospheres \citep{ciardi, absil_2006, defrere_2012, absil_2013, ertel_2014}, which has motivated the developments presented here. These precise observations may have deep implications on our understanding of planetary system evolution and direct detection of exoplanets in the habitable zone. Since NIR excesses are typically detected at the few sigma level, even modest improvements in the data analysis may have a noticeable impact on the measured exozodiacal light level and detection significance. The developments presented here are motivated by the search for exozodi phenomena with optical and near-infrared interferometry, but our results apply to high accuracy visibility observations in general. \\

The main interferometric observable with a two beam interferometer is the \emph{visibility} modulus, which is a measure of interference fringe contrast. A consequence of the Van Cittert-Zernike theorem is that the fringe visibility modulus is related to the angular radiance distribution of the astrophysical source by a Fourier transform, that goes from angular space to  \emph{baseline} space, where the baseline is defined as the projected telescopes separation. The visibility is measured by taking the frequency power spectrum of the fringes, and the area $S$ under the fringe frequency peak\footnote{Assuming that $I(t)$ is a dimensionless fringe pattern, its power spectrum ($|\tilde{I}(\omega)|^2$) has units of time squared, and the area under the fringe peak ($S=\int |\tilde{I}(\omega)|^2 d\omega $) has units of time. } can be related to the visibility as \citep{benson}

\begin{equation}
  |V|^2 = 4\Delta \sigma v S, \label{visibility}
\end{equation}

where $v$ is the fringe scanning speed in a co-axial beam-combiner, and $\Delta\sigma$ is related to the optical bandwidth $\Delta\lambda$ and the central wavelength $\lambda$ as $\Delta\sigma=\Delta\lambda/\lambda^2$. Typically, the measurable quantity for a fringe exposure is the squared visibility modulus rather than the visibility modulus. Detailed descriptions of the extraction of the visibility from an interferogram can be found in \citet{roddier_1984}, and \citet{vincent_drs} and \citet{kervella} discuss the case of a spatially filtered co-axial combiner. 

Ground-based interferometers are affected by atmospheric turbulence, which deforms the optical wave-front at time-scales that are typically shorter than $1\,\mathrm{s}$. To first order, atmospheric turbulence causes the fringe position to drift in time, which makes long exposures difficult and generally many ($\sim 100-200$) short ($\ll 1\,\mathrm{s}$) fringe exposures need to be taken. The uncalibrated (raw) visibility modulus is then estimated from a set of many squared visibility moduli. 

This paper discusses two main points: $i)$ how to estimate the uncalibrated (raw) visibility from an ensemble of fringe exposures while relaxing some assumptions about the statistical distribution of the measured raw visibilities, and $ii)$ how to calibrate the raw visibilities with a more general approach. We address the uncalibrated visibility estimation problem in Section \ref{median}.  The problem of data calibration is addressed in Section \ref{cal_sec}. In Section \ref{real_data} we test our strategies with real interferometric data.

\section{Estimating the uncalibrated visibility} \label{median}

The purpose of this section is to find a statistical estimator of the uncalibrated visibility with minimal bias and uncertainty, based on a sequence of $|V|^2$ measurements. Typical approaches to estimate the visibility from an ensemble of fringe exposures essentially fit the data to a particular statistical distribution. One approach to estimate the squared visibility is to simply take the ensemble mean and standard deviation of the squared visibilities.  However, the mean of $|V|^2$ is a biased estimator because the mean of squared-visibilities ($\langle |V|^2 \rangle$) is not equal to the square of the mean ($\langle |V| \rangle ^2$), the two quantities differing by the variance ($\sigma_{|v|}^2$). A classic approach has been to essentially average the variance subtracted squared visibilities \citep{tango_and_twiss, shao_1988}, assuming photon and detector noise. This approach has been successfully applied by interferometrists for measuring angular separations (e.g. \citet{npoi_1998}), diameters, and limb-darkening (e.g. \citet{npoi_2001}).

Given that the visibility modulus can only take positive values, another approach is to fit the data to a log-normal distribution \citep{chara}. However, if data do not follow the assumed distribution, the corresponding estimator may become biased and less suitable for highly accurate measurements.

A simple example where assuming Gaussian distributed data can yield a biased result is in the presence of outliers in an otherwise Gaussian distribution of visibilities. One example of such a probability distribution $\mathcal{P}$ can be approximately described as 

\begin{equation}
  \mathcal{P}(|V|) \approx (1-\epsilon) \mathcal{N}(\mu,\sigma) + \epsilon\,\delta(|V|-\mu-n\sigma),
\end{equation}

where $\mathcal{N}$ is a normal distribution of mean $\mu$ and standard deviation $\sigma$, and $\delta$ is a Dirac distribution weighed by a probability $\epsilon$ of the outlier to occur at $n$ standard deviations away from the mean $\mu$. In this example the bias induced by the outliers on the mean of $V$ is simply $\langle |V| \rangle-\mu = n\sigma\epsilon$, which is greater than a standard deviation when $n\epsilon>1$, and can become comparable to the effects we are trying to measure. A well-known method to alleviate this problem is to reject visibilities that lie far from the ensemble \emph{median}, which is less sensitive to outliers, but the rejection criteria are somewhat arbitrary and we would like to avoid rejecting data as much as possible.

\begin{figure}
  \begin{eqnarray}
  \includegraphics[scale=0.45]{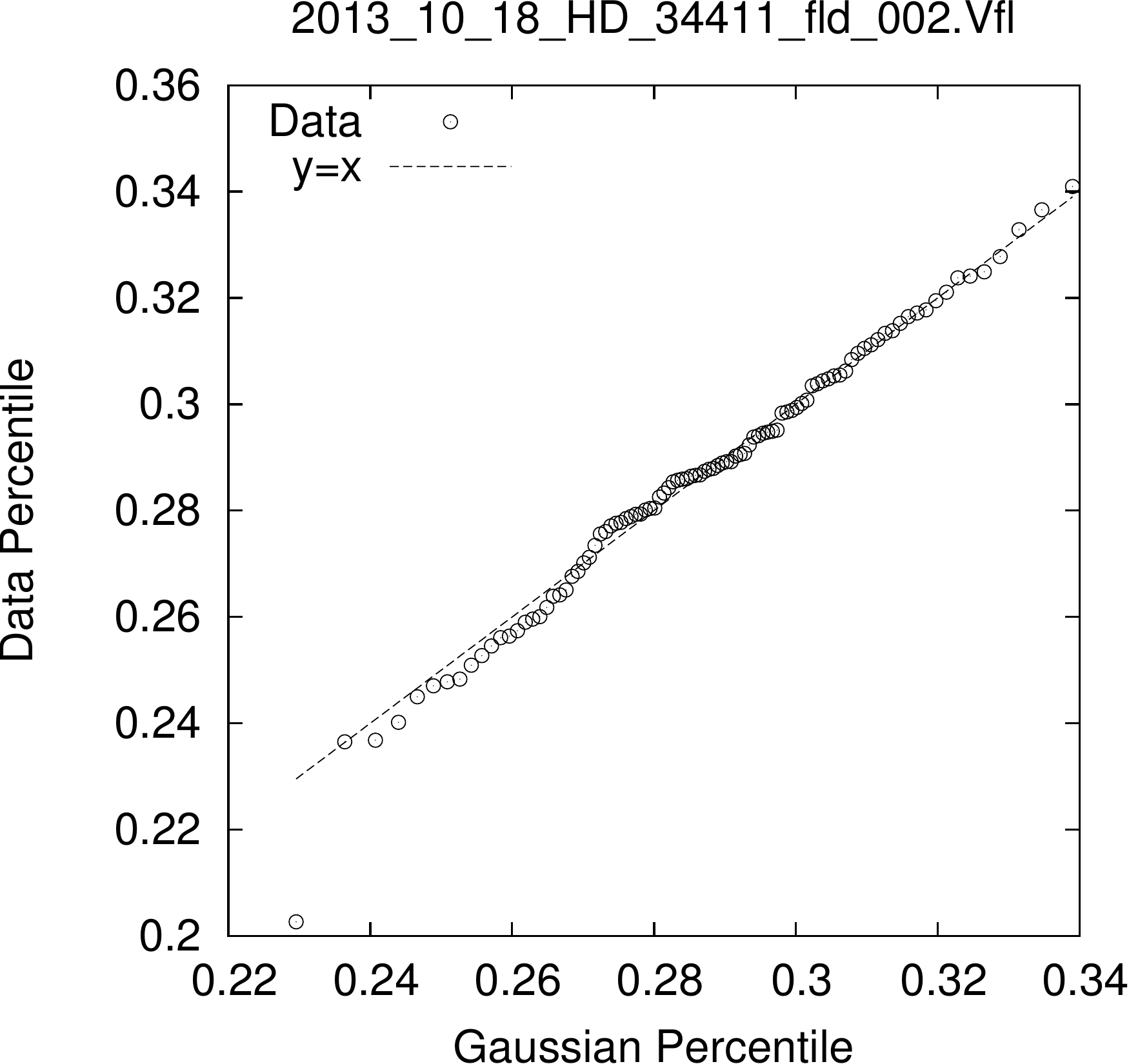} & 
  \includegraphics[scale=0.45]{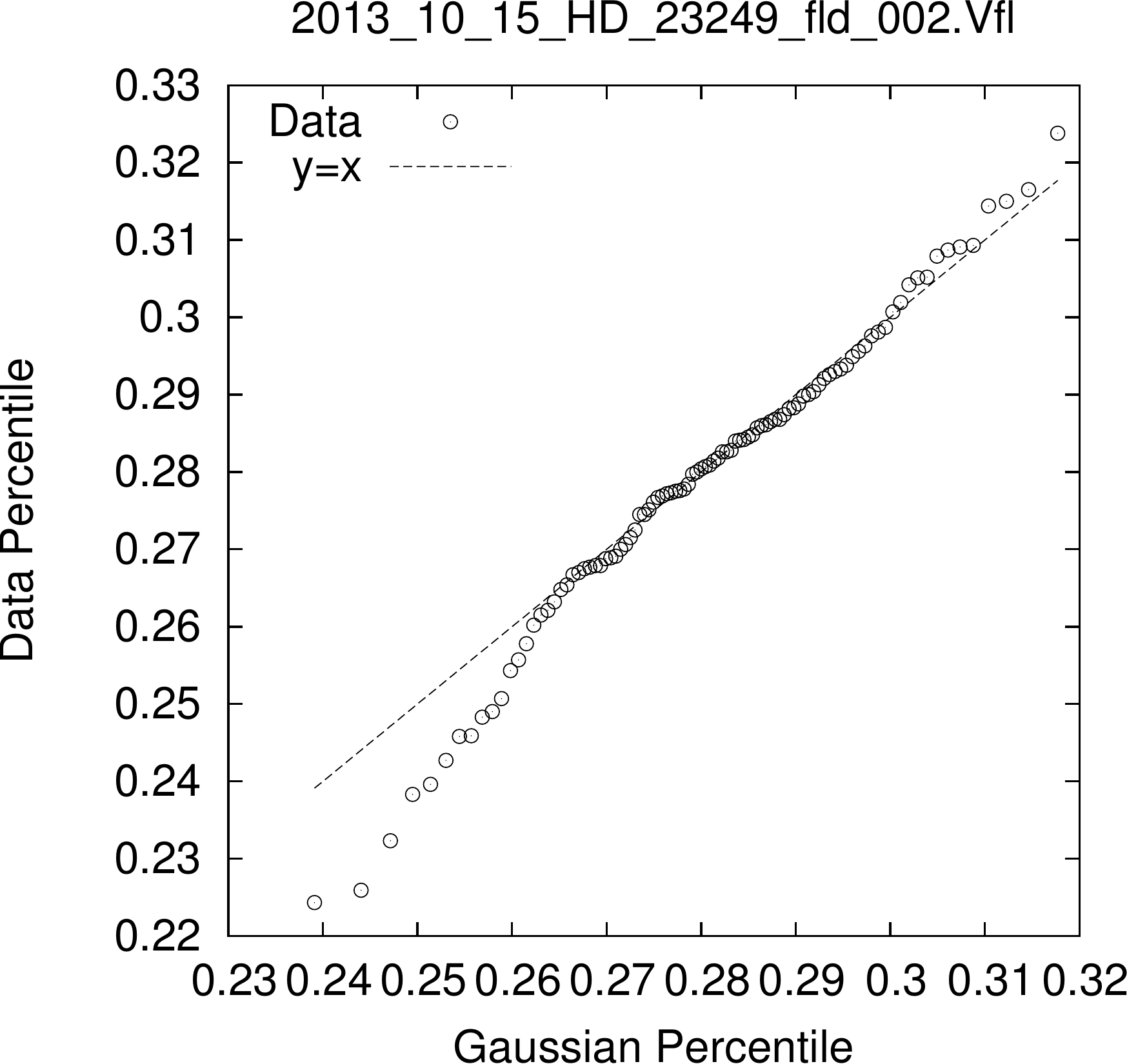} \nonumber
  \end{eqnarray}
  \caption{\label{sample_data} From an ensemble of CHARA measured stellar visibilities, or rather $\sqrt{|V|^2}$, for two different objects (HD 34411 on the left panel and HD 23249 on the right panel), we calculate the data percentiles shown on the vertical axis, and compare them to the percentile that would correspond to a best-fit Gauss distribution.}
\end{figure}

Another example of non-Gaussian data can be seen in Figure \ref{sample_data}, where we show probability plots for two different sets of visibilities obtained at the Center for High Angular Resolution Astronomy (CHARA) \citep{chara} using the  JouFLU beam-combiner (Jouvence of the Fiber-Linked Unit for Optical Recombination \citep{nic}). For each set of visibilities we compare the percentiles derived from the measured visibilities to those expected from a Gaussian distribution with the same mean and standard deviation as observed in the data. The left-most point of each plot is the first percentile, and the right-most point the 99$^{th}$ percentile. The left plot of Figure \ref{sample_data} shows data that are basically Gaussian, while the right plot shows data departing strongly from Gaussian statistics, since the distribution displays a heavy tail at low visibilities. In the case of Gauss distributed data (Fig. \ref{sample_data}, left), the median and the mean are virtually indistinguishable, while in the example of non-Gaussian data, the mean is nearly 2 standard deviations away from the ensemble median.

In general, the true distribution of the data is unknown, and assuming a particular distribution may lead to incorrect estimates of the visibility. A solution is to have an estimator that makes no assumptions on the statistical nature of the data. 

\subsection{The Median as a Visibility Estimator} \label{med_sec}

We propose to use the \emph{Median} as an estimator of the visibility, or rather 

\begin{equation}
V_{\rm{median}}= \sqrt{\rm{Median}(|V|^2)}, \label{median_eq}
\end{equation}

The median is known to be very resilient to outliers and seems to have the smallest bias among all the other visibility estimators we have experimented with as we will show below.

\begin{figure}
  \begin{center}
    \begin{eqnarray}
    \includegraphics[scale=0.32]{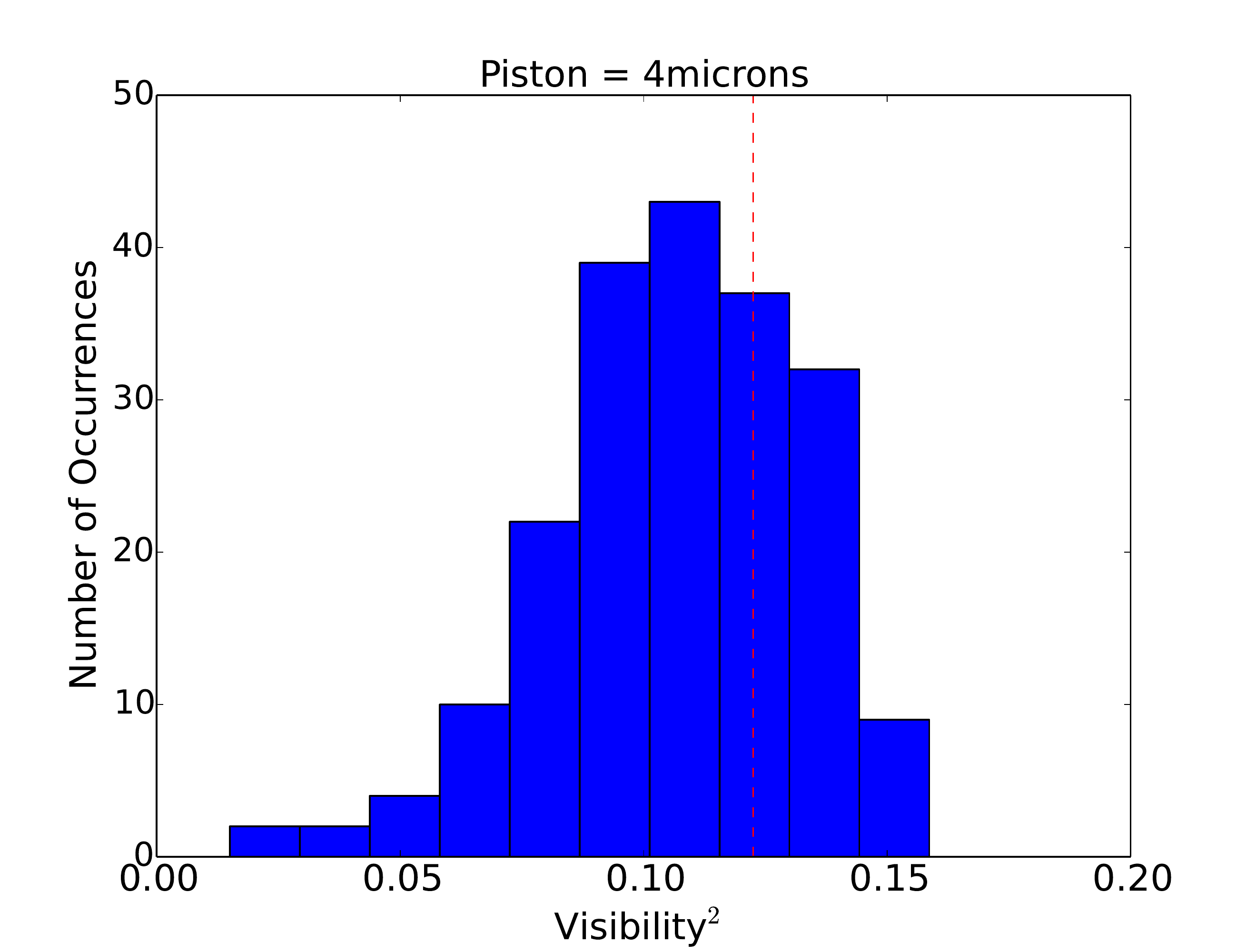} & \includegraphics[scale=0.32]{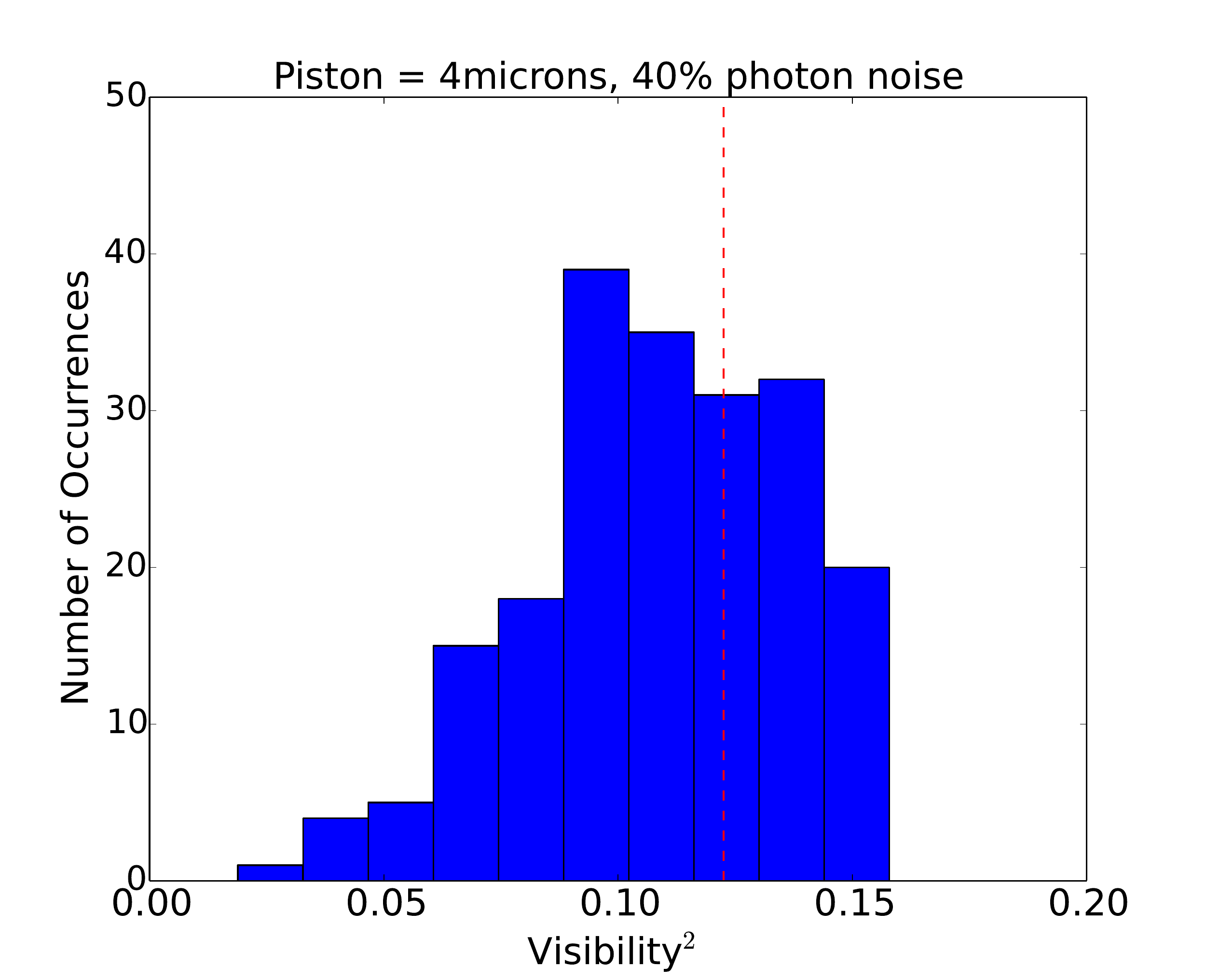} \nonumber \\
    \includegraphics[scale=0.32]{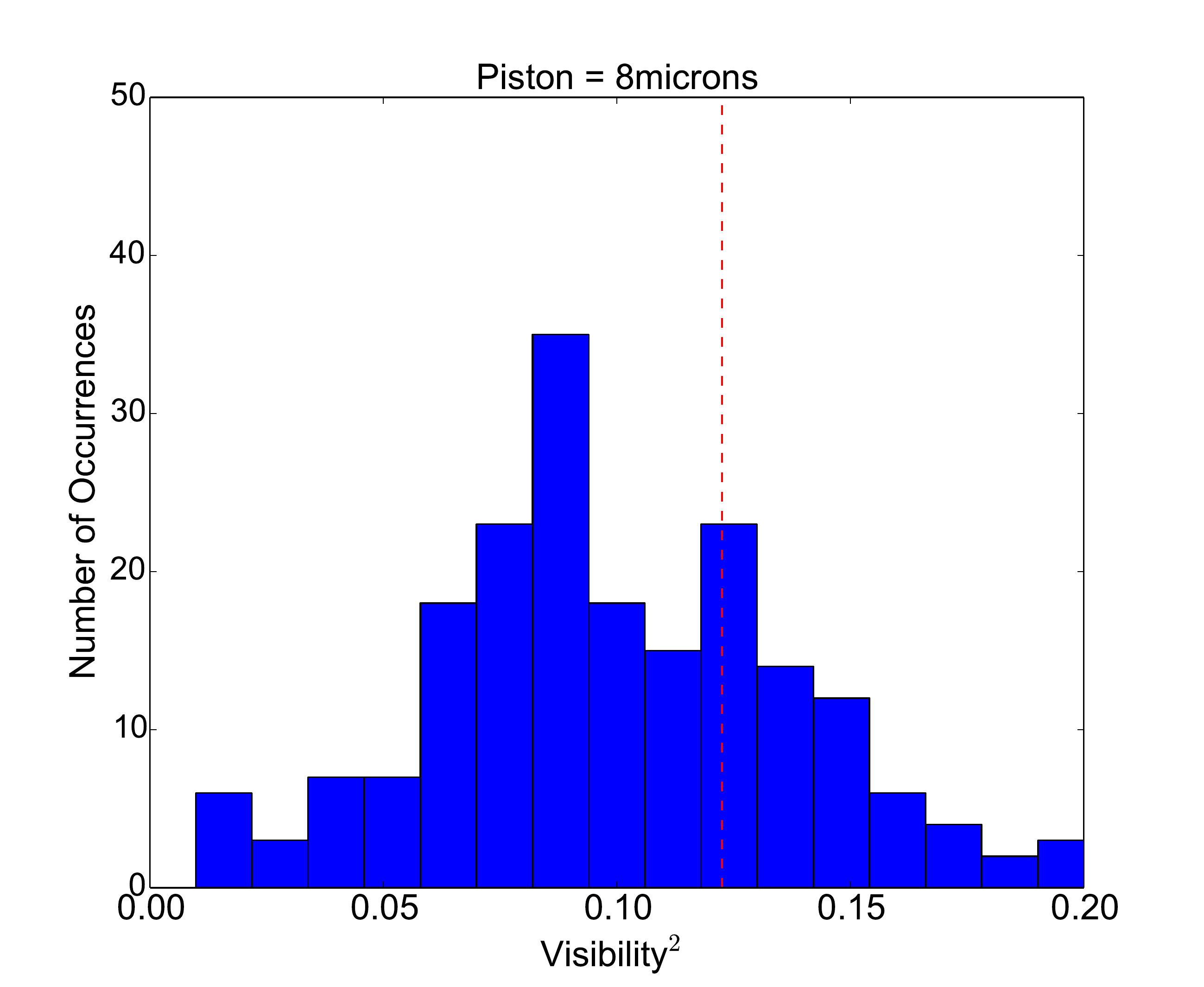} & \includegraphics[scale=0.32]{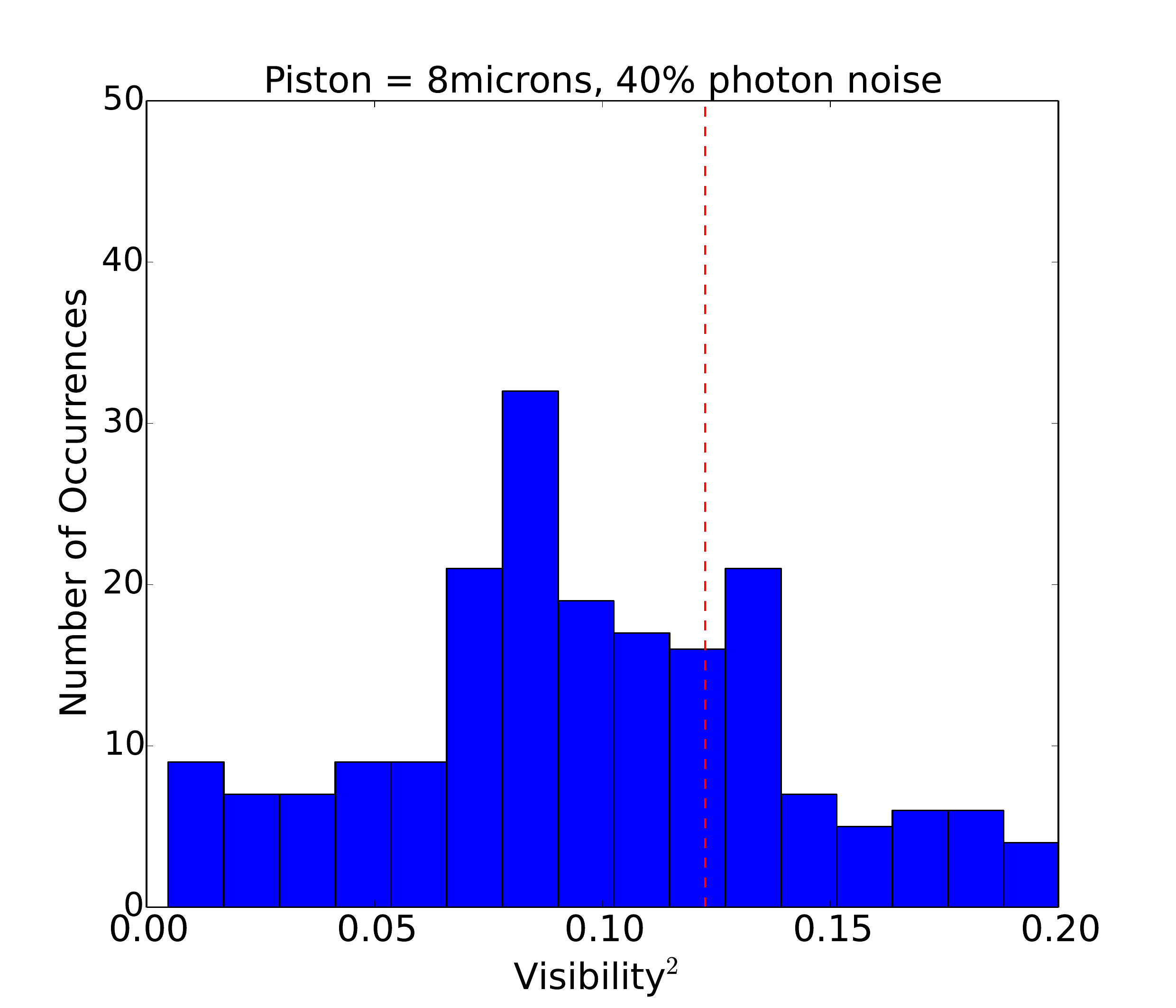} \nonumber 
    \end{eqnarray}
  \end{center}
  \caption{\label{piston_histogram} Histograms of simulated squared visibilities, with nominal $\mu=0.35$ (shown in red). The left row shows distributions only affected only by atmospheric piston with a Kolmogorov power spectral distribution. The atmospheric piston simulation assumes a wind speed of $10\,\mathrm{m/s}$, a $33\,\mathrm{m}$ baseline, $1\,\mathrm{m}$ apertures, and an OPD standard deviation of $4\,\mu\mathrm{m}$ (top panels) and $8\,\mu\mathrm{m}$ (bottom panels) for each scan of $0.5\,\mathrm{s}$ in duration (200 total scans for each histogram). For the case of the $4\,\mu \mathrm{m}$ rms OPD simulation, the visibility estimators presented in the appendix give $V_{median}=0.325\pm0.0032$, $V_{mean}=0.323\pm0.0034$, $V_{lognorm}=0.320\pm0.0031$, and $V_{norm}=0.320\pm 0.0032$. When the OPD approaches the JouFLU coherence length of $\sim 10\mu \mathrm{m}$, the visibility becomes difficult to estimate. The second column includes simulated photon (Poisson) noise added to each fringe exposure. }
\end{figure}

To investigate the performance of various estimators, we consider two main sources of noise on the squared visibility: \textit{i)} atmospheric wave-front distortions, which are dominated by differential-piston noise in the case of a spatially-filtered beam combiner, and \textit{ii)} detector photon noise and/or background noise. Piston noise can be modeled as multiplicative noise, and to illustrate the effect of differential-piston noise on the measured visibility, we first simulated interferograms affected by a time dependent Optical Path Difference (OPD) introduced by the atmosphere as described by \citet{perrin_1997} and \citet{choquet_2014}. The time dependent OPD was modeled as having a Kolmogorov Power Spectral Density, and a standard deviation of 4 microns during $0.5\,\mathrm{s}$ exposures. The effect of detector noise on the squared visibility estimate of Eq. 1, can be modeled as the difference of the square of two independent Gaussian variables, i.e. the difference between the real integrated detector-noise Power Spectral Density (PSD), and the estimated integrated detector-noise PSD. Then we computed the visibility modulus for each interferogram using Eq. \ref{visibility}, resulting in the histogram of squared visibilities shown in Figure \ref{piston_histogram}. The simulated piston distribution of $4\,\mu\mathrm{m}$ (rms) results in a standard deviation error of $\sim 0.003$ on the visibility. In order to further simplify our simulations, we model the squared visibility distribution as

\begin{equation}
    |V|^2=\mu^2\left(1 + \mathcal{N}_1\right)^2+\mathcal{N}_2, \label{model}
\end{equation}

where $\mu$ is the \emph{true} (positive) visibility, and $\mathcal{N}_1$, $\mathcal{N}_2$ are zero mean random noise variables with respective uncertainty $\sigma_1$ and $\sigma_2$. In this simplified model, $\mathcal{N}_1$ corresponds to atmospheric piston noise, and the term $\mathcal{N}_2$ results from the detector noise, where we assume that the power-spectrum bias has been removed (see \citet{perrin_2003} for a discussion on the power spectrum bias). Note that the mean of $|V|^2$ is a biased estimator because $\langle |V|^2 \rangle=\mu^2(1 + \sigma_1^2)$. Assuming a particular distribution for the visibilities (e.g. lognormal) will also result in a noticeable visibility bias because the data do not follow the assumed distribution perfectly. Below we describe some basic cases when the median is an unbiased estimator.

\begin{figure}
  \begin{eqnarray}
      \includegraphics[scale=0.43]{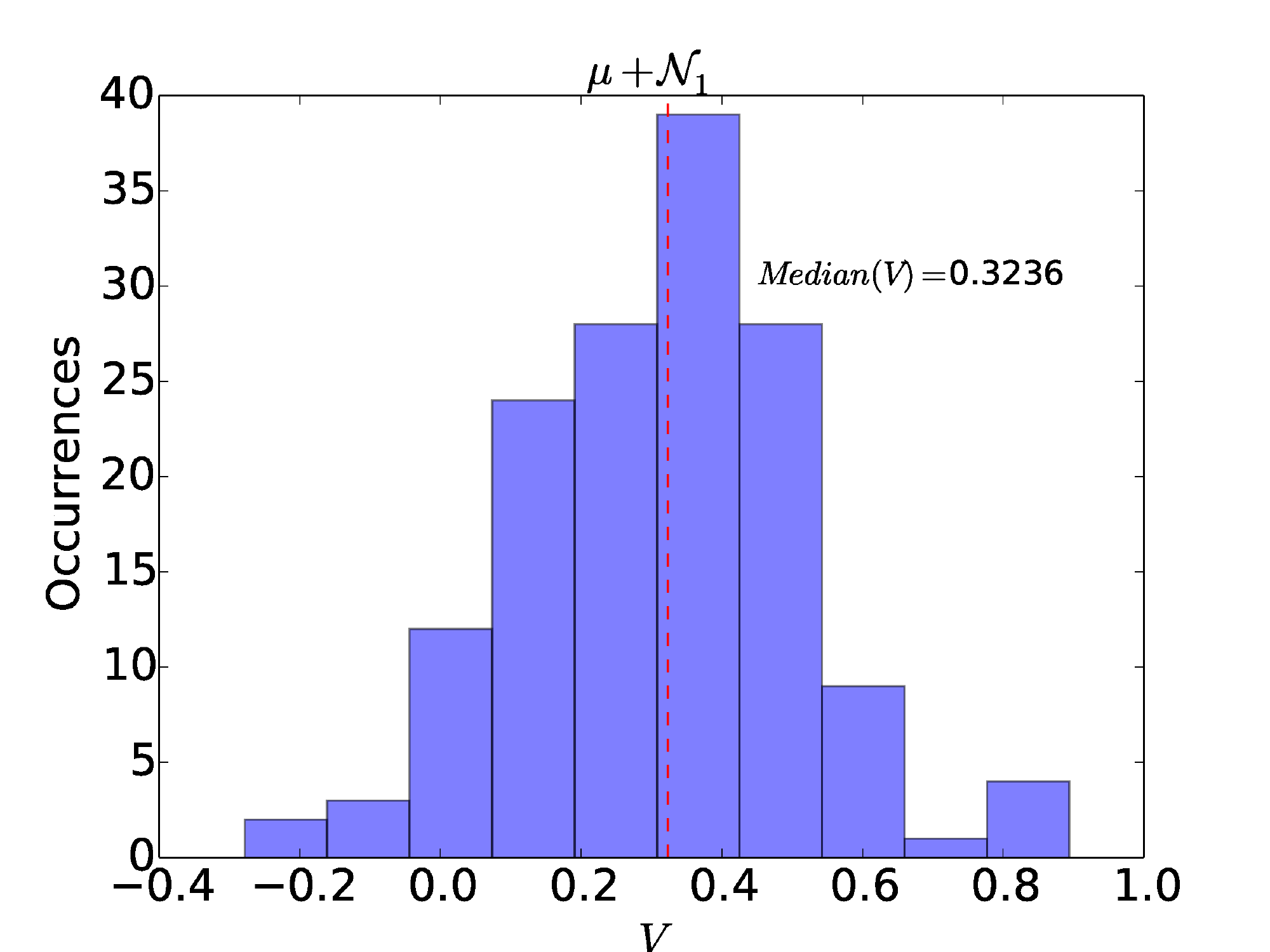} &
      \includegraphics[scale=0.43]{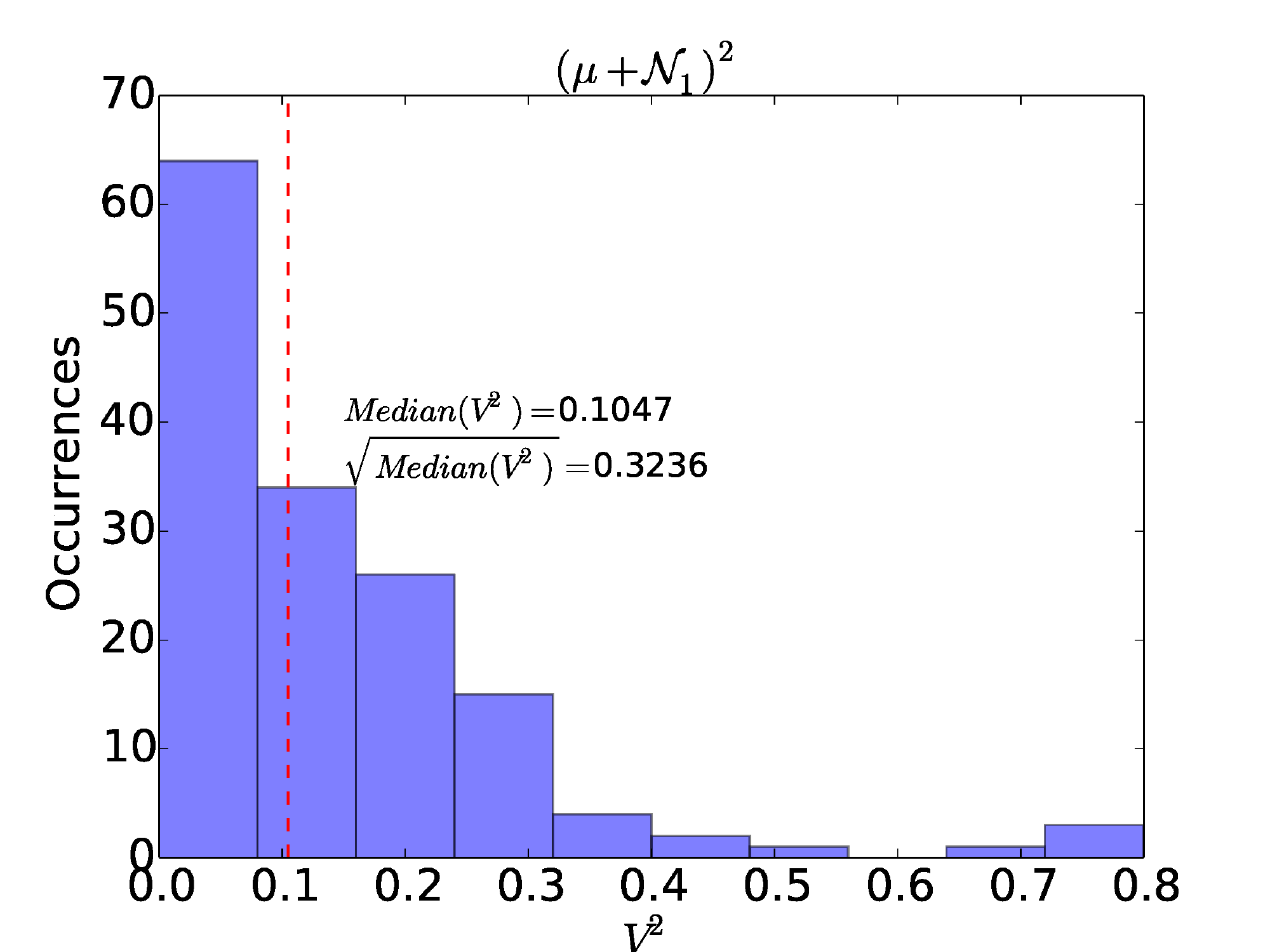} \nonumber 
  \end{eqnarray}
    \caption{\label{mu_plus_n1} Left panel: Simulated distribution of measured visibilities $|V|$ affected by multiplicative noise of the form $(1+\mathcal{N}_1)$, where $\mathcal{N}_1$ has a zero median. Note that visibilities can extend to negative values. Right panel: Corresponding distribution of the squared visibilities. In the case shown here and whenever  $|\rm{min}(|V|)|<\rm{Median(|V|)}$, we have simply $\sqrt{\rm{Median}(|V|^2)}=\rm{Median}(|V|)$. The median estimator is found to have the smaller bias of the 4 estimators, and a 1$\sigma$ estimation uncertainty consistent with that of the other estimators. Only the $V_{\rm{lognorm}}$ estimator provides a slightly smaller uncertainty, but at the expense of a much larger bias. }
\end{figure}

In the particular case when $\mathcal{N}_2\sim 0$ (bright star and piston-limited noise), and further assuming that $|\mu(1 + \rm{min}(\mathcal{N}_1))|<\mu(1+\rm{Median}(\mathcal{N}_1))$, a condition that is generally satisfied unless visibilities are very low and noisy (see Figure \ref{mu_plus_n1}), we have $\rm{Median}(|V|^2)=(\mu(1+\rm{Median}(\mathcal{N}_1)))^2$. This means that if the $\mathcal{N}_1$ noise has zero median, regardless of its statistical distribution, then the median of the observed $|V|^2$ can be used as an unbiased estimate of the object's visibility, i.e. $\sqrt{\rm{Median}(|V|^2)}=\mu$.  Similarly, in the opposite limiting case when $\mathcal{N}_1=0$ (photon-limited noise), we get $\rm{Median}(|V|^2)=\mu^2+\rm{Median}(\mathcal{N}_2)$. If we further assume that $\rm{Median}(\mathcal{N}_2)=0$, then $\rm{Median}(|V|^2)=\mu^2$ if $\rm{Median}(\mathcal{N}_2)=0$, regardless of $\mathcal{N}_2$'s statistical distribution. 

\begin{figure}
  \begin{eqnarray}
    \includegraphics[scale=0.42]{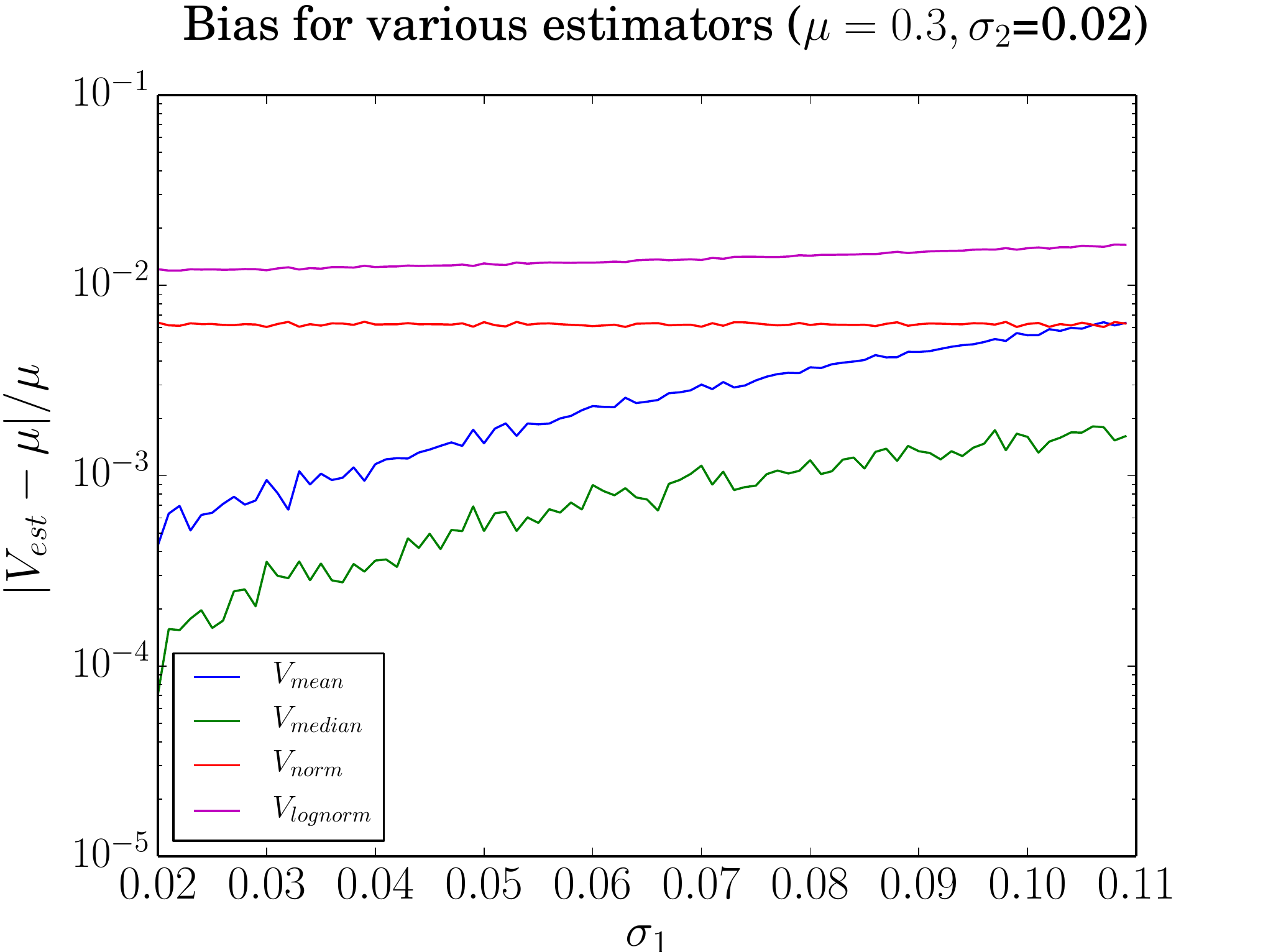} &
    \includegraphics[scale=0.4]{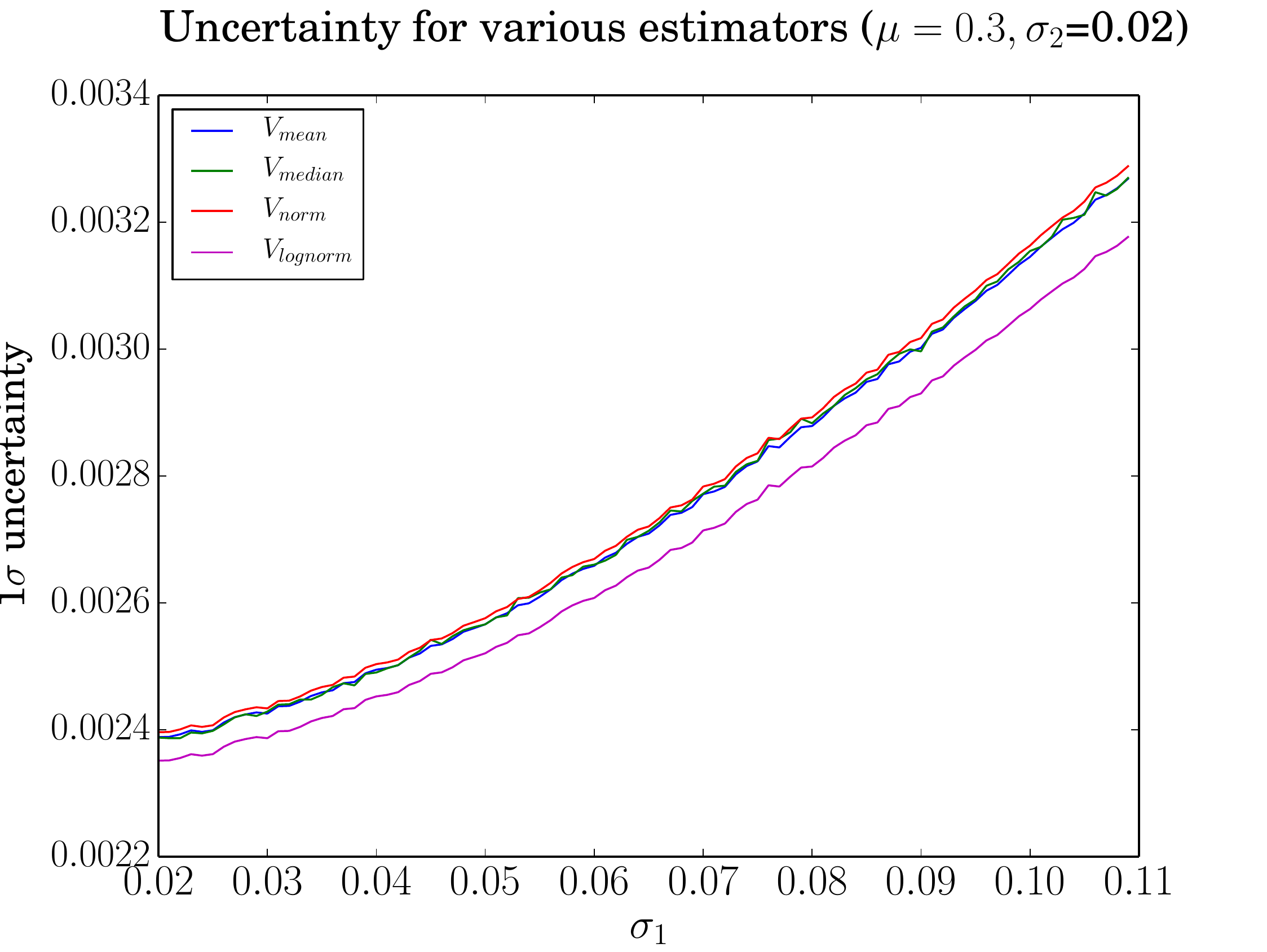}\nonumber\\
    \includegraphics[scale=0.42]{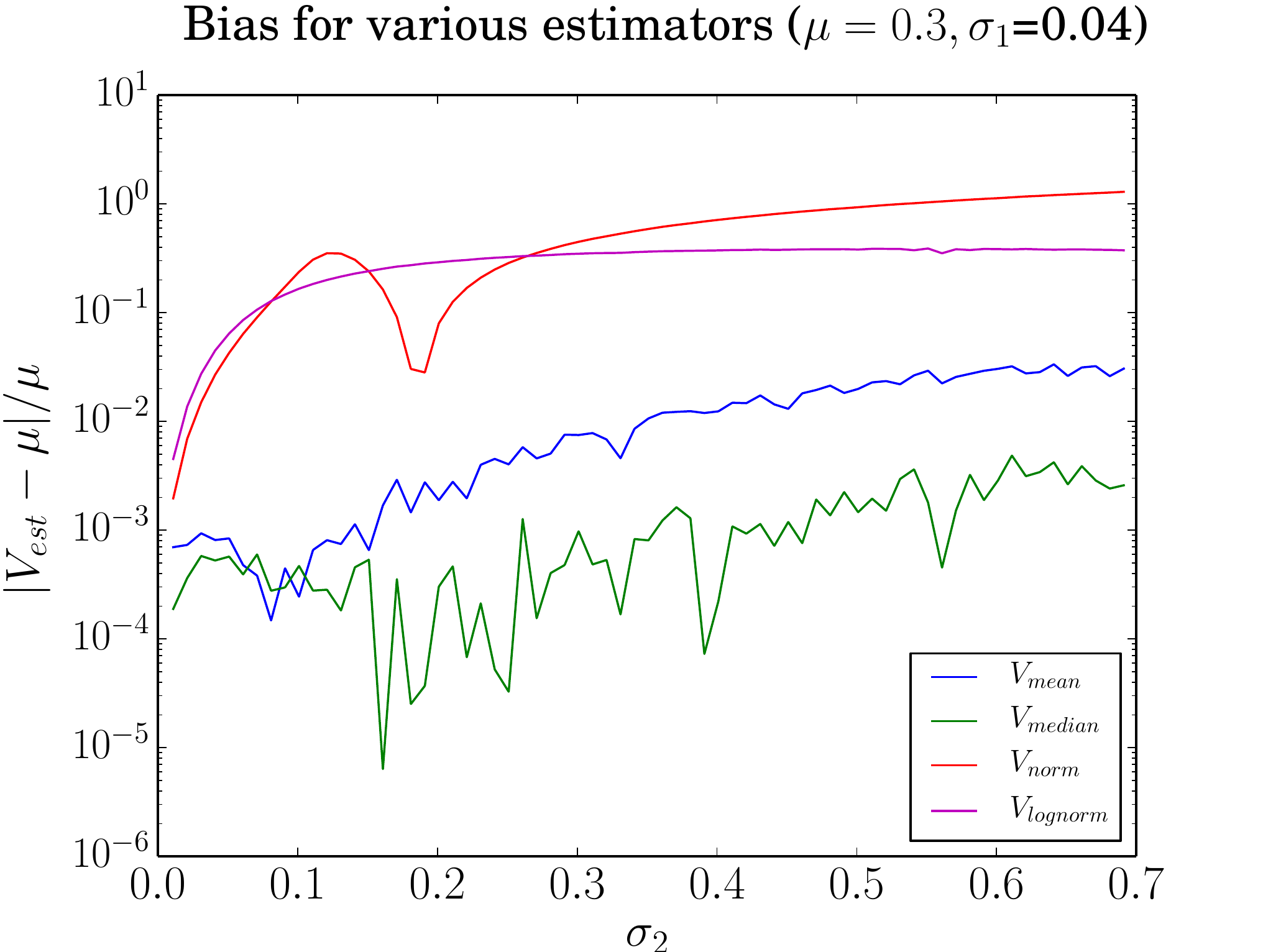} &
    \includegraphics[scale=0.42]{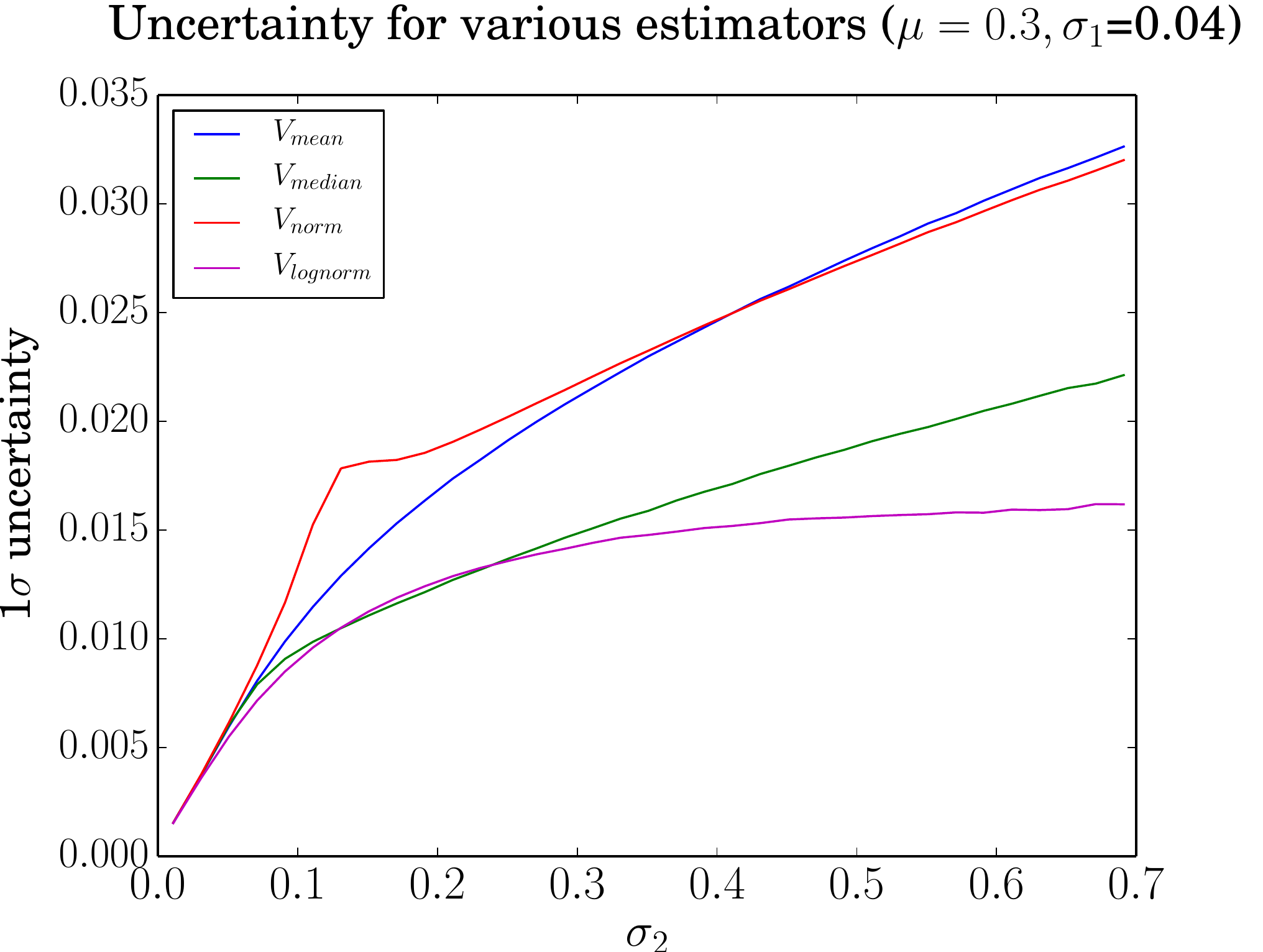}\nonumber
  \end{eqnarray}
  \caption{\label{bias}Using Equation \ref{model}, and assuming Gaussian noise for $\mathcal{N}_1$, and $\chi^2$ distributed noise for $\mathcal{N}_2$, we compare the relative biases expected for different estimators (left panels). The assumed visibility is $\mu=0.3$ and we take a fixed $\sigma_2=0.02$ for the top-left panel, which  corresponds to the effect of detector noise on a typical target star with K=3, and a fixed $\sigma_1=0.04$ (the piston noise, estimated from the simulation presented in Figure \ref{piston_histogram}) for the bottom-left panel. The right panels shows the corresponding $1\sigma$-spread for the various estimators based on an ensemble of 200 simulated scans. the range of $\sigma_2$ values corresponds to a range in stellar magnitudes K=0-4.9.}
\end{figure}

In general we are somewhere in between the two extreme cases discussed above, so we simulated distributions of 200 visibility points using Eq. \ref{model}, taking $\mathcal{N}_1$ as a Gaussian random variable, and $\mathcal{N}_2$ as a $\chi^2$ distributed variable, with various realistic values of $\sigma_1$, $\sigma_2$. Then we computed the estimation bias resulting from different visibility estimators. In all cases, the relative bias is defined as $|V_{\rm{est}}-\mu|/\mu$, where $V_{\rm{est}}$ is one the 4 different estimators described in \citet{theo_2005}, and also provided in Appendix A of this manuscript, namely: $V_{\rm{lognorm}}$, $V_{\rm{norm}}$, $V_{\rm{median}}$ (defined in equation \ref{median_eq} above), and $V_{\rm{mean}}$ (a simple mean and standard deviation). In Figure \ref{bias} (left) we compare the bias, and find that in the context of the model described by equation \ref{model}, $V_{\rm{median}}$ generally has the smallest bias among the estimators tested.

From the bottom-left panel of Figure \ref{bias}, we note that the visibility bias, using the median estimator, remains below $\sim 0.5\%$, even for the faintest stars ($\sigma_2=10^{-4}-0.7$, corresponding to stellar magnitudes ranging between K=0-4.9). This shows that using calibrators of very similar brightness to the science target is not required when using the median estimator, and for stars brighter than $K\approx 4.9$. In any case, in the JouFLU survey we adopt a conservative approach, and use calibrators that do not differ by more than $\sim1$ magnitude from the science target. At some point ($\sigma_2\sim 0.3$), the uncertainties become too large for highly precise visibility measurements, as shown in the bottom-right panel of Figure \ref{bias}. This is actually close to the point where the detector noise becomes comparable to the signal counts. The JouFLU beam combiner typically detects $\sim 100$ coherent photons per fringe exposure for an unresolved  $\mathrm{K}\approx 3$ star, while the background (dark) counts rms is $\sim 10$ counts per fringe exposure. Since the $\mathcal{N}_2$ term standard deviation results from the difference between two squared Gaussians of standard deviation  $\sim 10/100$, the resulting standard deviation for a typical K=3 star is $\sigma_2\approx 2\times(10/100)^2 = 0.02$. Therefore, detector noise is the limiting source of noise when the stellar magnitude approaches $K \sim 4.5$ ($\sigma_2 \sim 0.3$).

Next we investigate the uncertainty of the median, which we estimate from bootstrapping the ensemble of visibility measurements resulting from individual fringe exposures as follows: from a set of $N=150$ exposures, a ``bootstrap sample'' of the data is generated by randomly selecting $N$ visibilities. Then $10^4$ such bootstrapped samples are generated and a median is computed for each of them. Last, we estimate the $68\%$ confidence interval by finding the $16^{th}$ and $84^{th}$ percentiles of the ensemble of bootstrap medians. We perform analogous simulations for the other 3 estimators and compare the uncertainties derived for all 4 estimators. As shown in Figure \ref{bias} (right), our simulations show that the $V_{\rm{median}}$ estimator has a very similar uncertainty to the $V_{\rm{mean}}$ and $V_{\rm{norm}}$ estimators, and is marginally larger than the uncertainty of $V_{\rm{lognorm}}$, which has the largest bias. In Section \ref{real_data} we discuss the performance of $V_{\rm{median}}$ with real interferometric data.

\section{Hybrid Calibration} \label{cal_sec}

The raw visibility $\mu$ must be calibrated in order to relate it to the angular brightness distribution of the science target. Optical interferometry requires calibrator stars whose angular diameters are known with sufficient precision  predict their expected theoretical visibility $V_{\rm{exp}}$ with high accuracy. The science target calibrated visibility can be expressed as

\begin{equation}
  V=\mu \, T,
\end{equation}

where $T$ is known as the transfer function, and defined in terms of the expected visibility of the calibrator $V_{\rm{exp}}$ and the calibrator's measured raw visibility $\mu_c$:

\begin{equation}
  T\equiv\frac{V_{\rm{exp}}}{\mu_c}.
\end{equation}

Calibrators are typically observed before and after the science target since the transfer function may vary in time, and we find the visibility transfer functions $T_1$ and $T_2$ at times $t_1$ and $t_2$. The problem we address here is how to best estimate the transfer function $T$ at the time of observation $t$ ($t_1<t<t_2$), which reduces to finding the weights $\alpha_1$ and $\alpha_2$ for the transfer functions $T_1$ and $T_2$, i.e.

\begin{equation}
  V=\mu\,T=\mu \left\{ \alpha_1 T_1 +\alpha_2 T_2 \right\}\frac{1}{\alpha_1+\alpha_2} 
  \label{calibrated_v}
\end{equation}

One possibility to find the weights $\alpha_1$ and $\alpha_2$ is to assume that the transfer function varies linearly in time, and perform a linear time-interpolation to estimate the transfer function at the time of observation. In this case the weights are 

\begin{equation}
  \alpha_1(t)=\left(\frac{t_2-t}{t_1-t_2}\right);\,\,\,  \alpha_2(t)=\left(\frac{t-t_1}{t_1-t_2}\right).
\end{equation}

Alternatively -and especially if the $T_1$ and $T_2$ estimated values differ by less than their individual error bars- it is also reasonable to assume that the transfer function did not change between $t_1$ and $t_2$, in which case we could calculate a ``weighted mean'' with statistical weights

\begin{equation}
  \alpha_1=\frac{1}{\sigma_1^2} ;\,\,\, \alpha_2=\frac{1}{\sigma_2^2}.
\end{equation}

We propose a solution for the coefficients ($\alpha_1$, $\alpha_2$) combining weights from the linear time interpolation and from the weighted mean:

\begin{eqnarray}
  \boxed{  \alpha_1(t)=\frac{1}{\sigma_1^2}\left(\frac{t_2-t}{t_1-t_2}\right)} & \boxed{\alpha_2(t)=\frac{1}{\sigma_2^2}\left(\frac{t-t_1}{t_1-t_2}\right) }.
\end{eqnarray}

Note that when $\sigma_1=\sigma_2$, equation \ref{calibrated_v} reduces to a linear time-interpolation, and when $t=(t_2+t_1)/2$, equation \ref{calibrated_v} reduces to a weighted mean. If $t$ is much closer to $t_1$ than to $t_2$, then $T_1$ will have a much larger weight than $T_2$, unless the uncertainty on $T_2$ ($\sigma_2$) is much smaller than that on $T_1$ ($\sigma_1$).

For computing the final uncertainty in the calibrated visibility (Eq. \ref{calibrated_v}) we first take into account the statistical uncertainties\footnote{In the case of the JouFLU beam combiner, the statistical uncertainties of each transfer function include the effects of correlations between the two interferometric channels as described by \citet{perrin_2003}} of $T_1$ and $T_2$.  The statistical uncertainty $\sigma_{\rm{stat}}$ on the estimated calibrated visibility comes from propagating the uncertainties on $\mu$, $T_1$ and $T_2$, and adding them in quadrature, i.e.:

\begin{eqnarray}
    \sigma_{\rm{stat}}^2 &=& \left(\frac{\partial V}{\partial \mu}\sigma_{\mu}\right)^2+
                       \left(\frac{\partial V}{\partial T_1}\sigma_1\right)^2+
                       \left(\frac{\partial V}{\partial T_2}\sigma_2\right)^2. \nonumber \\
                      &=& T^2 \sigma_{\mu}^2 + \left( (\alpha_1\sigma_1)^2 + (\alpha_2\sigma_2)^2\right)\frac{\mu^2}{ (\alpha_1+\alpha_2)^2}
\end{eqnarray}

In addition we also consider the systematic uncertainty coming from the departure of $T_{1}$ from $T_{2}$. This systematic uncertainty is calculated as a weighed standard deviation, i.e.

\begin{equation}
  \sigma_{\rm{syst}}^2=\mu^2\frac{\left(T_1-T \right)^2\alpha_1 + \left( T_2-T \right)^2\alpha_2}{\alpha_1+\alpha_2}. \label{sigma_syst}
\end{equation}

The final uncertainty in the visibility can be estimated by adding errors in quadrature as

\begin{equation}
  \sigma_{|V|}^2=\sigma_{\rm{stat}}^2+\sigma_{\rm{syst}}^2.
\end{equation}

According to Eq. \ref{sigma_syst}, the distribution of the systematic uncertainty is a $\chi^2$ distribution, which can result in a high probability for the over-estimation of $\sigma_{syst}$. The results presented above can be generalized to an arbitrary number of calibrators, or alternatively we could resort to Gaussian process estimation (as suggested by the referee), but we restrict the results presented below to the nearest neighbor approximation, i.e. two calibration measurements: one before and one after the science target. 

\section{Testing on Real Data} \label{real_data}
\subsection{All Stars}

We tested the performance of the different estimators and calibration methods on a set of $166$ calibrated visibility points obtained with JouFLU/CHARA between 2013 and 2015 for the exozodiacal light survey, for stars as bright as K=1.4 and as faint as K=4.7. These tests were only performed on stars for which the naked star model produced a $\chi^2$ that is not too high (reduced $\chi^2$ lower than 3.0 with $\sim$5 degrees of freedom) and no excess was detected (with a threshold of $3\sigma$) in the star + dust model. This reduces possible  biases due to the presence of exozodiacal structures or stellar companions around the target stars, and still includes  the majority of targets in the overall survey sample ($\sim 70\%$). The  goal here is to see which estimator and calibration method gives the smallest bias and uncertainty, while the full results of the exozodi survey will be presented in an upcoming paper. In order to measure the bias, we compare each calibrated visibility point to a modeled photospheric visibility based on long baseline angular diameter measurements obtained by \citet{taby1} with CHARA/CLASSIC. If long-baseline measurements are not available, we use a surface brightness (V-K) angular diameter to model the visibility, which is acceptable for the JouFLU program since stars are mostly unresolved at the short ($\sim 30\,\mathrm{m}$) baselines used, and angular diameter uncertainties induce negligible error at these short baselines. For example a star with a uniform-disk angular diameter of $(1\pm0.05)\,\mathrm{mas}$ would introduce an uncertainty of $\sim 0.07\%$ in the interferometric visibility at the $33\,\mathrm{m}$ baseline). 

\begin{figure*}
  \begin{eqnarray}
    \includegraphics[scale=0.4]{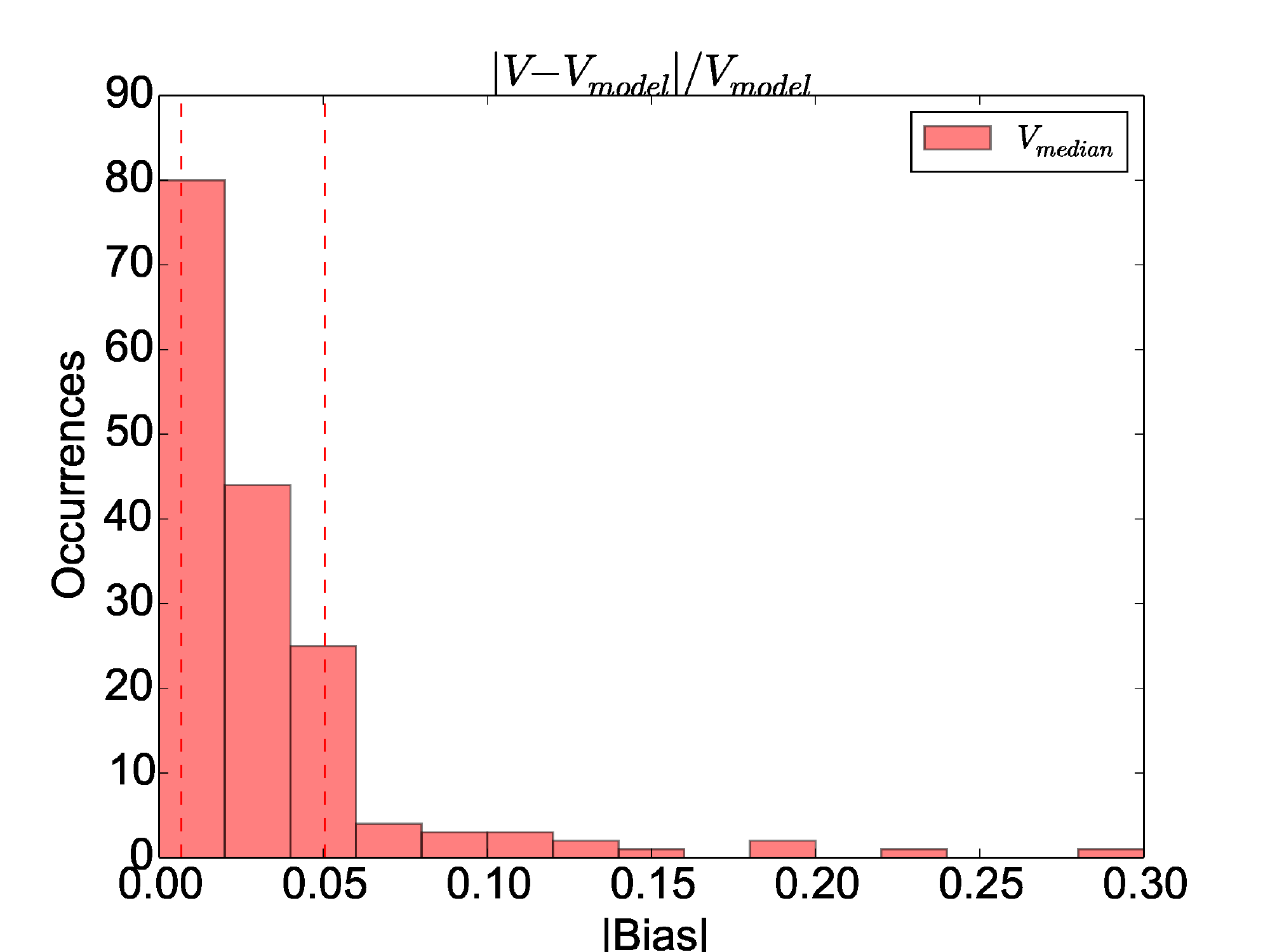} & \hspace{1.0cm} \includegraphics[scale=0.4]{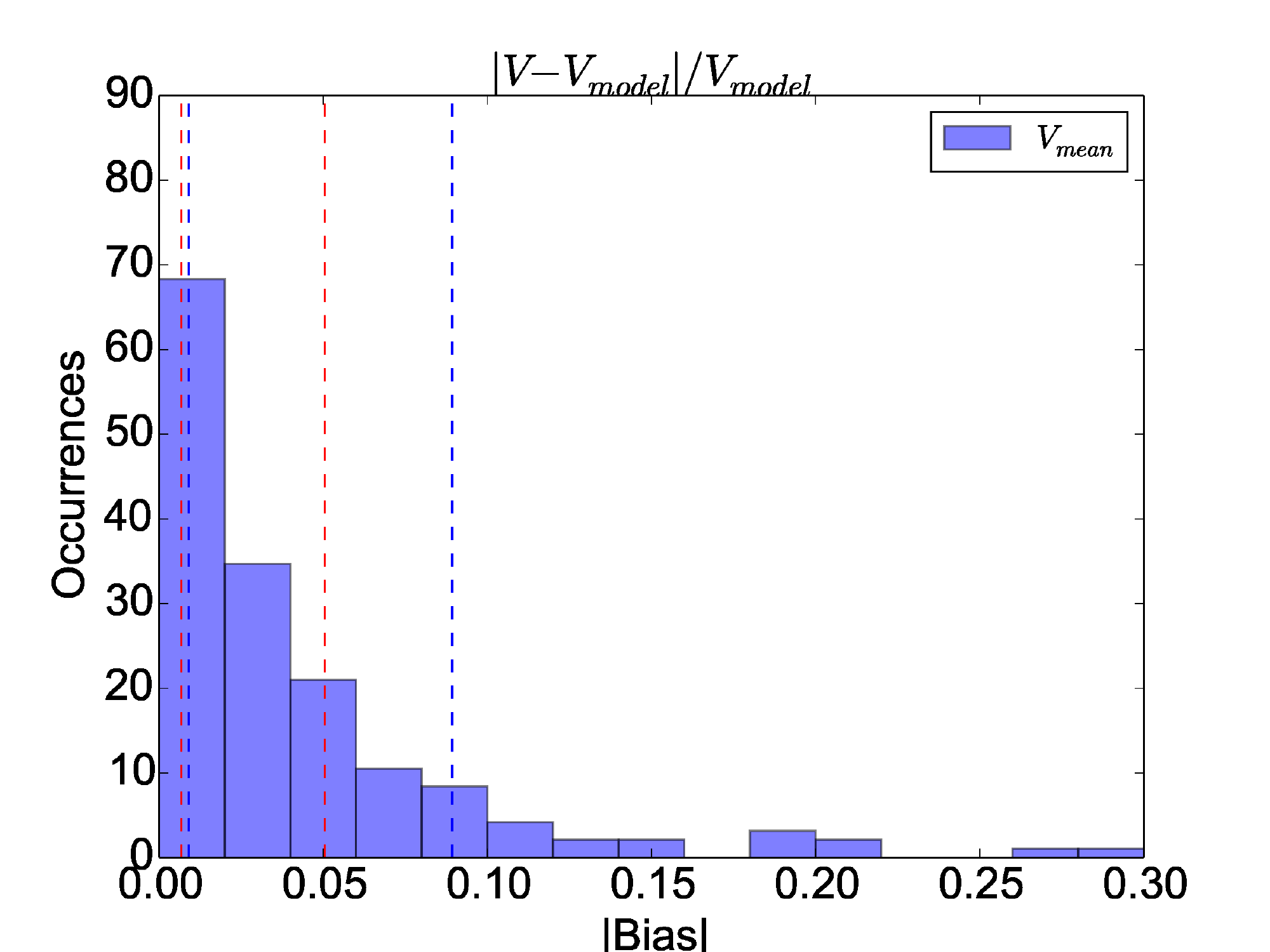}\nonumber\\ 
    \includegraphics[scale=0.4]{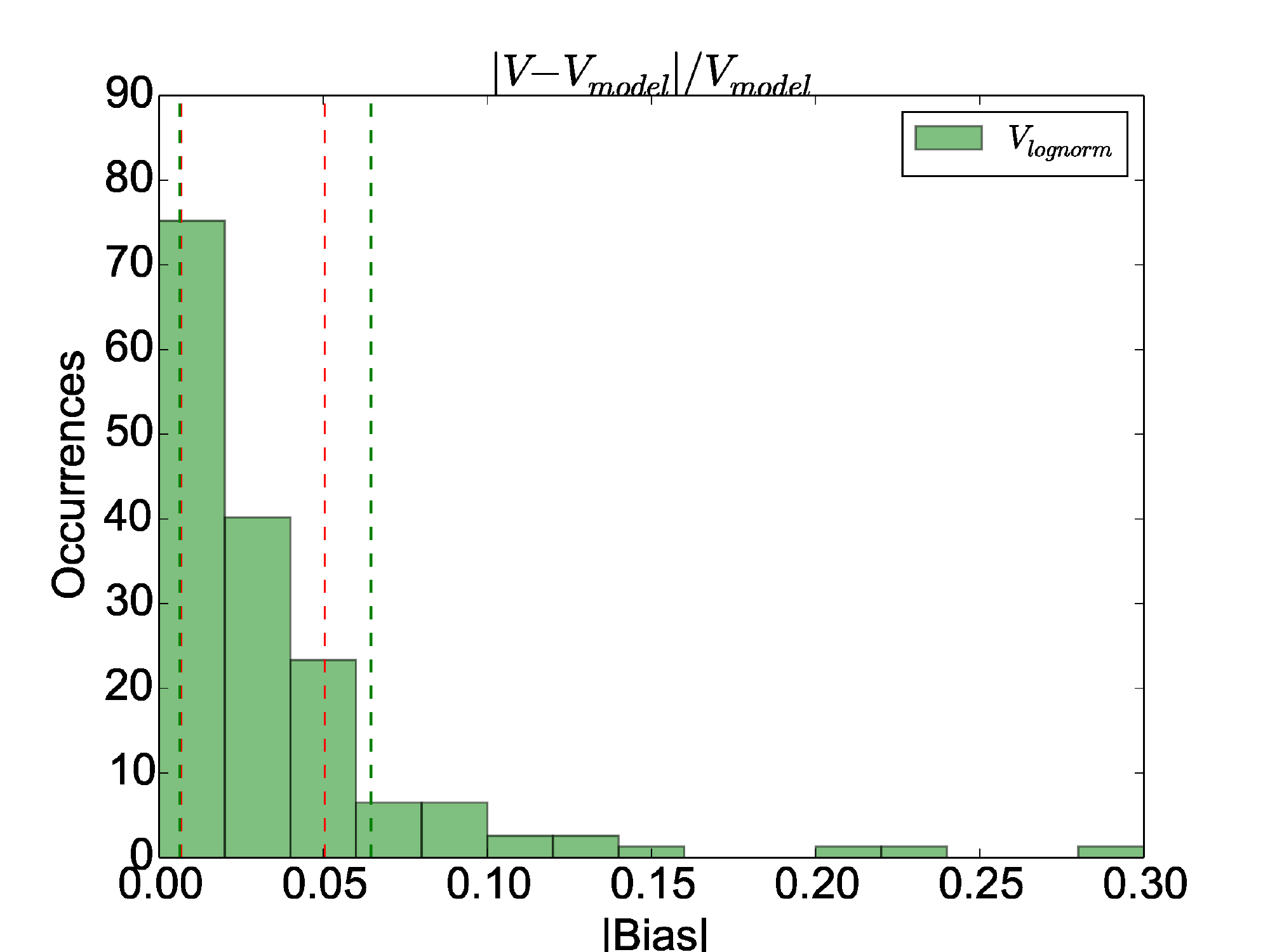} & \hspace{1.0cm}\includegraphics[scale=0.4]{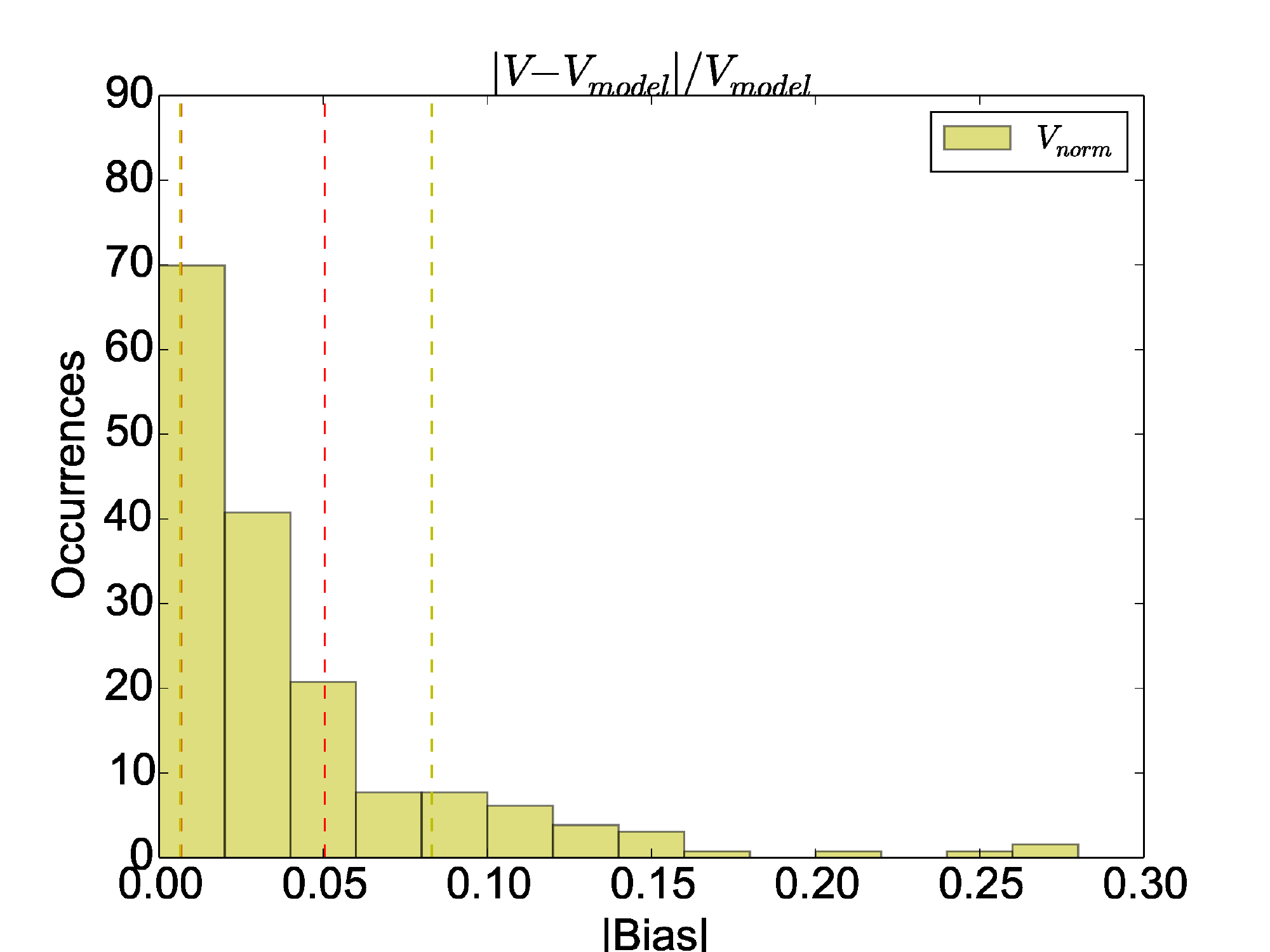} \nonumber 
  \end{eqnarray}
  \caption{\label{abs_bias}Histogram of the absolute value of the relative bias for various estimators. The vertical dotted lines represent the $68\%$ confidence interval. The confidence interval of $V_{\rm{median}}$ (shown in red) is smaller than for all other estimators considered here ($V_{\rm{mean}}$, $V_{\rm{lognorm}}$, and $V_{\rm{norm}}$). }
\end{figure*}


We compare the performance of various estimators ($V_{\rm{median}}$, $V_{\rm{mean}}$, $V_{\rm{norm}}$, $V_{\rm{lognorm}}$) and calibration methods (LI: Linear time Interpolation, and HI: Hybrid Interpolation discussed in Sec. \ref{cal_sec}), i.e. we analyzed the whole data set using 8 different visibility estimation methods. Figure \ref{abs_bias} shows histograms of $|V-V_{\rm{model}}|/V_{\rm{model}}$,  Table \ref{summary} provides the median values of the $|$bias$|$, the median visibility error, and also shows the uncertainties ($68\%$ CI) associated to these median values, which are calculated via bootstrapping techniques. Figure \ref{abs_bias} and Table \ref{summary} show that when using the $V_{\rm{median}}$ estimator with the Hybrid interpolation calibration, hereafter $V_{\rm{median,HI}}$, the biases and uncertainties are generally lower. We also find that their statistical dispersion is lower, i.e, that the range of observed biases and visibility uncertainties is smaller. In other words, the $V_{\rm{median,HI}}$ method provides more robust and more consistent calibration results.  

\begin{figure}
  \begin{eqnarray}
    \hspace{-0.5cm}\includegraphics[scale=0.345]{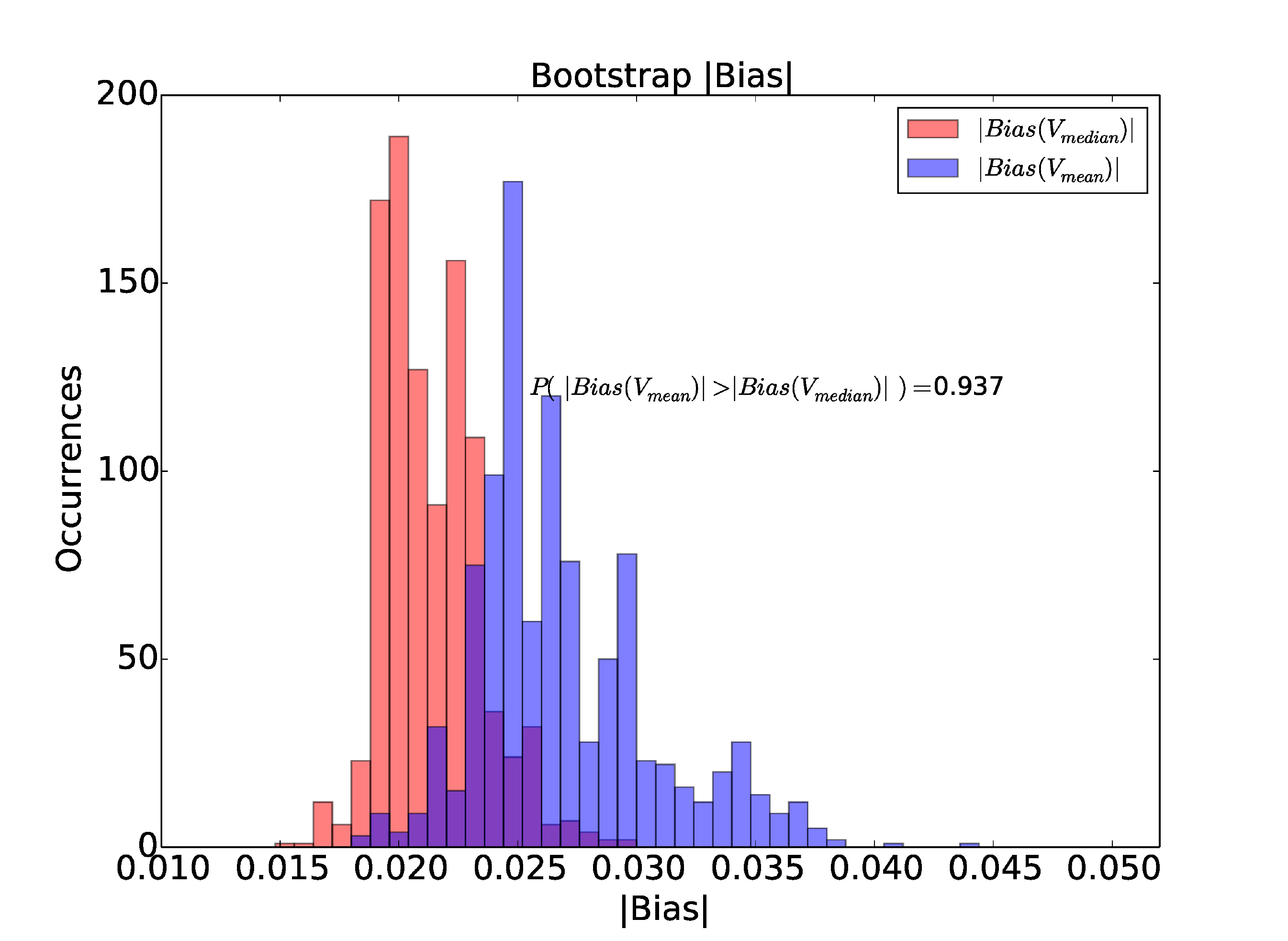} &
    \hspace{-1.28cm}\includegraphics[scale=0.345]{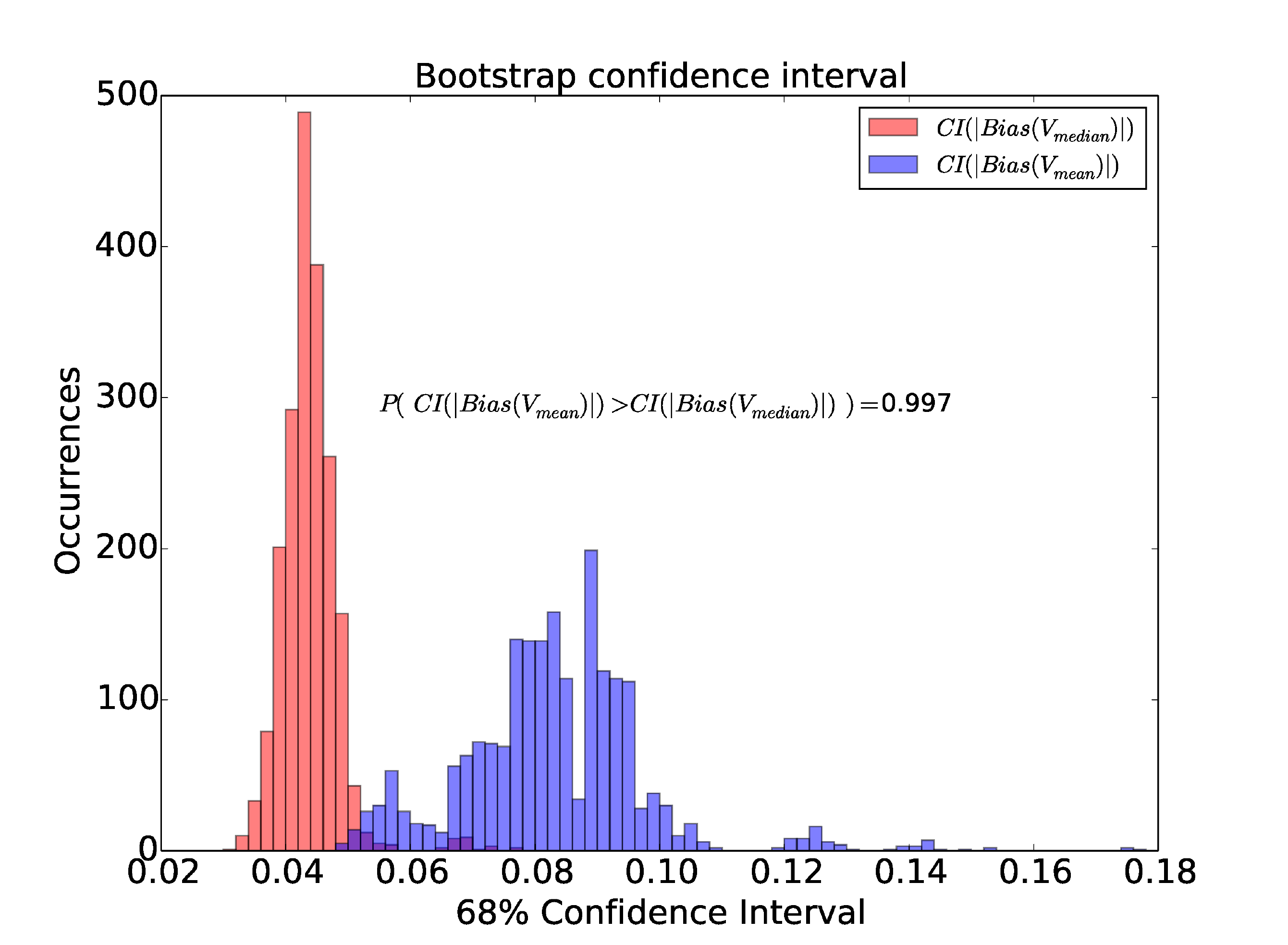} \nonumber
  \end{eqnarray}
  \caption{\label{bootstrap} Bootstrap values for the $|\rm{Bias}|$ (left panel) and its $68\%$ CI for the $V_{\rm{median}}$ and $V_{\rm{mean}}$ estimators (right panel). The probability that $|\rm{Bias}|$ is greater for $V_{\rm{mean,LI}}$ than for $V_{\rm{median}}$ is $94\%$, and the probability that the Confidence Interval of $|\rm{Bias}|$ is greater for $V_{\rm{mean,LI}}$ than for $V_{\rm{median}}$ is $99\%$. The distributions show that $V_{\rm{median}}$ has a smaller bias and error than $V_{\rm{mean}}$.}
\end{figure}

To quantify the statistical significance of our findings, we use the following bootstrapping method: for each visibility estimation method, there is a set of 166 calibrated visibility measurements, which we resample many (1000) times to form many bootstrapped sets. For each bootstrapped set we calculate the $84^{th}$ and $16^{th}$ percentiles of the bias and visibility error. Then, to calculate the probability (confidence level CL) that $V_{\rm{median,HI}}$ is a better estimator, we simply count the number of times that the $68\%$ confidence interval (CI) is smaller for the $V_{\rm{median,HI}}$ estimator. Figure \ref{bootstrap} illustrates this for the bias and its uncertainty found for the $V_{\rm{median}}$ and $V_{\rm{mean}}$ estimators.

\begin{table}
\caption{\label{summary} In the second column we report the median relative percent $|$bias$|$ (defined as $|V-V_{\rm{model}}|/V_{\rm{model}}$) and its bootstrapped $68\%$ confidence level. The last column shows the error for each visibility estimation method. We provide values for various estimators and interpolation methods (LI: Linear time Interpolation, HI: Hybrid Interpolation). }

  \begin{center}
    Visibility Bias and Error\\
    \begin{tabular}{|l|l|l|}
      \hline
      Estimator     &   \small{$|V\!\!-\!V_{\rm{model}}|/V_{\rm{model}}$}  & Median Error \\ \hline
      $V_{\rm{median,HI}}$ & $(2.11 \pm 0.19)\%$ & $(2.1 \pm 0.14)\%$ \\ \hline
      $V_{\rm{median,LI}}$  & $(2.26 \pm 0.22)\%$ &  $(2.2 \pm 0.16)\%$\\ \hline
      $V_{\rm{lognorm,HI}}$ & $(2.32 \pm 0.28)\%$  & $(2.2 \pm 0.16)\%$ \\ \hline
      $V_{\rm{lognorm,LI}}$ & $(2.75 \pm 0.34)\%$ & $(2.3 \pm 0.17)\%$ \\ \hline
      $V_{\rm{mean,HI}}$    & $(2.48 \pm 0.28)\%$ &  $(2.6 \pm 0.19)\%$ \\ \hline
      $V_{\rm{mean,LI}}$  & $(2.60 \pm 0.32)\%$ & $(2.7 \pm 0.18)\%$ \\ \hline
      $V_{\rm{norm,HI}}$ & $(2.29 \pm 0.29)\%$ & $(2.8 \pm 0.15)\%$  \\ \hline
      $V_{\rm{norm,LI}}$ &  $(2.49 \pm 0.27)\%$ & $(2.8 \pm 0.16)\%$ \\ \hline
    \end{tabular}    
  \end{center}

\end{table}

As shown in Table \ref{bias_prob} (third column), the most statistically significant improvement ($>95\%$ CL) is in the reduction of the confidence intervals of $|V-V_{\rm{model}}|/V_{\rm{model}}$. Table \ref{bias_prob} shows that most of the improvement comes from using the median estimator described in Sec. \ref{med_sec}, and a slight improvement is due to the Hybrid interpolation method described in Sec. \ref{cal_sec}. 

\begin{table}
  \caption{\label{bias_prob} Probability that the $|$Bias$|$ of $V_{\rm{median,HI}}$ is reduced. The second column shows the probability that $\rm{Median}|V-V_{\rm{model}}|/V_{\rm{model}})$ is smaller for the $V_{\rm{median,HI}}$ estimator than for the other estimators listed in the first column. The third column shows the probability, $P(CI (|V_{\rm{median,HI}}-V_{\rm{model}}|/V_{\rm{model}})$, that the $68\%$ confidence interval of $|V_{\rm{median,HI}}-V_{\rm{model}}|/V_{\rm{model}}$ is smaller when $V_{\rm{median,HI}}$ is used.}
  \begin{center}
    Probability of bias reduction for $V_{\rm{median}}$\\
  \begin{tabular}{|l|c|c|}
    \hline 
    Estimator &  $P(|B_{\rm{median,HI}}|)<$  & $P(CI |B_{\rm{median,HI}}|)<$ \\ \hline 
    $V_{\rm{mean,LI}}$      &             0.94                  & 0.99                          \\ \hline 
    $V_{\rm{mean,HI}}$       &            0.86                  & 0.99                          \\ \hline 
    $V_{\rm{norm,LI}}$       &            0.86                  & 0.98                          \\ \hline 
    $V_{\rm{norm,HI}}$       &            0.67                  & 0.97                          \\ \hline 
    $V_{\rm{lognorm,LI}}$    &            0.70                   & 0.96                          \\ \hline
    $V_{\rm{lognorm,HI}}$    &            0.56                   & 0.80                          \\ \hline
    $V_{\rm{median,LI}}$    &             0.53                   & 0.68                          \\ \hline
  \end{tabular}
  \end{center}
\end{table}

Similarly, we quantify the reduction of the visibility uncertainty resulting from the use of $V_{\rm{median,HI}}$. As shown in Table \ref{visibility_error}, we find that the median uncertainty is significantly ($95\%$ CL) smaller by as much as $\sim 20\%$ relative to the $V_{\rm{norm}}$ and $V_{\rm{mean}}$ estimators, and comparable to the $V_{\rm{lognorm}}$ and $V_{\rm{median,LI}}$. As for the confidence interval of the visibility error, we find that $V_{\rm{median,HI}}$ does significantly better than most estimators, but not significantly better than $V_{\rm{lognorm,HI}}$ and $V_{\rm{median,LI}}$.

\begin{table}
  \caption{\label{visibility_error}Probability that $V_{\rm{median,HI}}$ has a smaller error than the estimators shown in the first column. The second column is the probability that the median visibility error is smaller for $V_{\rm{median,HI}}$ than for the other visibility estimation methods. The third column is the probability that the $68\%$ confidence interval of the visibility error is smaller for the $V_{\rm{median,HI}}$}
  \begin{center}
    Probability of error reduction for $V_{\rm{median}}$\\
  \begin{tabular}{|l|c|c|}
    \hline
        Estimator & \small{$P( Med(\sigma_{\rm{median,HI}}))<$}  &  \small{$P(CI(\sigma_{\rm{median,HI}}))<$}\normalsize \\ \hline
        $V_{\rm{mean,LI}}$           & 0.97                       &  1.0                        \\ \hline
        $V_{\rm{mean,HI}}$           & 0.96                        & 0.99                       \\ \hline
        $V_{\rm{norm,LI}}$           &  1.0                       &  0.97                    \\ \hline
        $V_{\rm{norm,HI}}$           & 1.0                        &   0.95                      \\ \hline    
        $V_{\rm{lognorm,LI}}$         & 0.75                    &   0.95                   \\ \hline
        $V_{\rm{lognorm,HI}}$         & 0.53                      &  0.78                     \\ \hline
        $V_{\rm{median,LI}}$          & 0.65                       &  0.63                     \\ \hline            
  \end{tabular}
  \end{center}
\end{table}

From the results of Table \ref{summary}, we also note that the median (typical) bias is close to the typical visibility uncertainty. This indicates that there is a global agreement of the visibility measurements with the stellar models for all visibility estimation methods. However, we consider $V_{\rm{median,HI}}$ a more robust estimator in view of the reduced dispersion of the bias and error.

\subsection{Influence of Stellar Brightness}

\begin{figure}
  \begin{center}
    \includegraphics[scale=0.35]{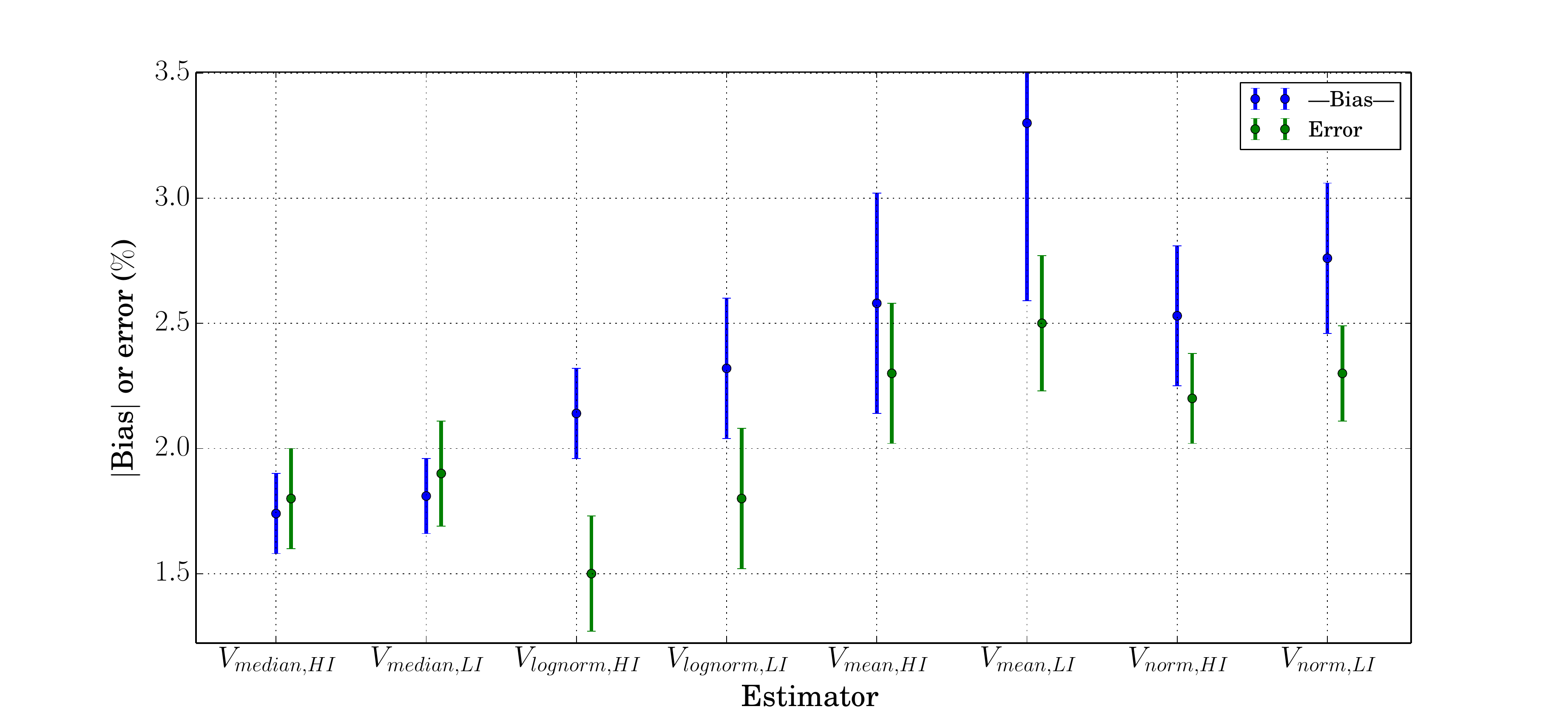}
  \end{center}
  \caption{\label{table_figure} Graphical representation of Table \ref{summary_bright}. The plot shows the abs(Bias) and the error for different estimators for stars with $K<3.5$. The median estimator has a smaller $|$bias$|$ than all the other estimators, and has a smaller error than smaller error than the other estimators except $V_{\rm{lognorm}}$. The statistical significance of our findings is shown in Table \ref{bias_prob_bright}.}
\end{figure}

Finally, we investigate the improvements as a function of stellar brightness, so we split our data set in two groups of comparable size: data points corresponding to brighter stars ($K<3.5$), and data corresponding to fainter stars ($K>3.5$). For the brighter stars, we find that the $|$bias$|$ (actually the relative bias absolute value) is $(1.74\pm 0.16)\%$ (for $V_{\rm{median,HI}}$), while for the group of fainter stars the relative $|$bias$|$ is $(2.3\pm 0.39)\%$. For the brighter stars, we find that the use of $V_{\rm{median,HI}}$ results in a significanlty smaller $\rm{Median}(|V-V_{\rm{model}}|/V_{\rm{model}})$ compared to most other estimation methods. Table \ref{summary_bright} and Figure \ref{table_figure} show that the (median) $|$bias$|$ is reduced by at least 23\% when the median estimator ($V_{\rm{median}}$)  is used, and Table \ref{bias_prob_bright} shows that these improvements are statistically significant at the $95\%$ CL. Compared to what we find with the whole stellar sample, we find no additional reduction in the visibility uncertainty for the bright stars as shown in Table \ref{summary_bright} and \ref{bias_prob_bright}. For the fainter stars, the performance of $V_{\rm{median,HI}}$ is still superior with probabilities comparable to those described in Tables \ref{bias_prob} and \ref{visibility_error}.

In general, special care must be taken for the calibration of the brightest stars, since the calibrator must be small enough to be considered a point-source, but of comparable brightness to the science target so that the interference fringes have similar noise characteristics. These two requirements are not generally met, since bright stars typically have larger angular diameters, and generally a compromise is made between brightness, closeness in the sky, and angular diameter \citep{boden_2007, merand_2005}. However, in the Exozodi survey, we have mainly used the smallest baseline of the CHARA array ($33\,\mathrm{m}$), for which most stars remain virtually unresolved, and with transfer functions that have a weak dependence on the stellar model. For example, the largest and brightest (K=2) calibrator used in our survey has a uniform angular diameter estimated at $1.71 \pm 0.024\,\mathrm{mas}$ (\citet{merand_2005} catalog). This corresponds to an interferometric visibility of $0.979\pm 0.0006$ at the $33\,\mathrm{m}$ baseline. Even assuming a pessimistic diameter uncertainty of 5\% on this worst case calibrator, the resulting visibility uncertainty is 0.7\%. Additionally, our results never rely on a single calibrator star, and we nominally use 3 different calibrator stars for each science target. A more extended discussion of our calibrators selection will be presented in our main survey results summary paper.
 
\begin{table}
  \caption{\label{summary_bright} Similar to Table \ref{summary}, but only showing results for stars with $K<3.5$.}
  \begin{center}
    Visibility Bias and Error\\

    \begin{tabular}{|l|l|l|}
      \hline
      Estimator          & \small{$|V\!\!-\!V_{\rm{model}}|/V_{\rm{model}}$}  & Median Error  \\ \hline
      $V_{\rm{median,HI}}$ & $(1.74 \pm 0.16)\%$ & $(1.8 \pm 0.19)\%$  \\ \hline
      $V_{\rm{median,LI}}$   & $(1.81 \pm 0.15)\%$ &  $(1.9 \pm 0.20)\%$ \\ \hline
      $V_{\rm{lognorm,HI}}$ & $(2.14 \pm 0.18)\%$  & $(1.5 \pm 0.21)\%$ \\ \hline
      $V_{\rm{lognorm,LI}}$ & $(2.32 \pm 0.28)\%$  & $(1.8 \pm 0.23)\%$ \\ \hline
      $V_{\rm{mean,HI}}$  & $(2.58 \pm 0.44)\%$  &  $(2.3 \pm 0.28)\%$ \\ \hline
      $V_{\rm{mean,LI}}$    & $(3.30 \pm 0.71)\%$  & $(2.5 \pm 0.27)\%$\\ \hline
      $V_{\rm{norm,HI}}$    & $(2.53 \pm 0.28)\%$ & $(2.2 \pm 0.18)\%$ \\ \hline
      $V_{\rm{norm,LI}}$    & $(2.76 \pm 0.30)\%$ & $(2.3 \pm 0.19)\%$ \\ \hline
    \end{tabular}    

  \end{center}
\end{table}

\begin{table}
  \caption{\label{bias_prob_bright} Similar to Table \ref{bias_prob}, but only showing results for stars with $K<3.5$.}
  \begin{center}
    Probability of bias reduction for $V_{\rm{median}}$\\
  \begin{tabular}{|l|c|c|}
    \hline 
    Estimator &  $P(|B_{\rm{median,HI}}|)<$  & $P(CI |B_{\rm{median,HI}}|)<$ \\ \hline 
    $V_{\rm{mean,LI}}$      &             1.0                  & 1.0                          \\ \hline 
    $V_{\rm{mean,HI}}$       &            0.99                  & 0.99                          \\ \hline 
    $V_{\rm{norm,LI}}$       &            0.99                  & 0.97                          \\ \hline 
    $V_{\rm{norm,HI}}$       &            0.99                  & 0.95                          \\ \hline 
    $V_{\rm{lognorm,LI}}$    &            0.99                   & 0.26                          \\ \hline
    $V_{\rm{lognorm,HI}}$    &            0.96                   & 0.70                          \\ \hline 
    $V_{\rm{median,LI}}$    &             0.61                   & 0.38                          \\ \hline
  \end{tabular}
  \end{center}
\end{table}

\begin{table}
  \caption{\label{visibility_error_bright} Similar to Table \ref{visibility_error_bright}, but only showing results for stars with $K<3.5$.}
  \begin{center}
    Probability of error reduction for $V_{\rm{median}}$
  \begin{tabular}{|l|c|c|}
    \hline
        Estimator & \small{$P( Med(\sigma_{median,HI}))<$}  &  \small{$P(CI(\sigma_{median,HI}))<$}\normalsize \\ \hline
        $V_{\rm{mean,LI}}$           & 0.99                       &  0.99                        \\ \hline
        $V_{\rm{mean,HI}}$           & 0.96                        & 0.97                       \\ \hline
        $V_{\rm{norm,LI}}$           &  0.99                       &  0.86                    \\ \hline
        $V_{\rm{norm,HI}}$           & 1.0                        &   0.99                      \\ \hline    
        $V_{\rm{lognorm,LI}}$         & 0.57                    &   0.95                   \\ \hline
        $V_{\rm{lognorm,HI}}$         & 0.62                      &  0.32                     \\ \hline
        $V_{\rm{median,LI}}$          & 0.59                       &  0.65                     \\ \hline            
  \end{tabular}
  \end{center}
\end{table}

\section{Discussion and conclusions}

The goal of this paper was to present the data analysis strategy for the CHARA/JouFLU exozodi survey. Our approach was to relax some of the assumptions made when estimating the visibilities, and to propose a more general methodology that is not only strictly valid for data ($V$ or $|V|^2$) affected by Gaussian errors, or errors with a priori perfectly known statistical distributions. We have shown that assuming a particular statistical distribution for an ensemble of fringe exposures may lead to a biased estimation of the uncalibrated visibility. Using the median as the visibility estimator is a natural choice since it is more resilient to outliers and skewed distributions. We also show that the median estimator is an unbiased estimator in several limiting cases, as long as the dominant source of measurement noise has a zero median, and whatever its statistical distribution is otherwise. Bootstrapping to find the errorbars of the median estimator also makes no assumptions on the statistics and is therefore quite general. Our proposed method for estimating the transfer function is also more general since it reduces to the commonly used linear interpolation when the calibrator's visibilities have similar uncertainties, but gives more weight to calibrators that have smaller uncertainties. 

We have performed tests with simulated and real data, and have concluded that the formalism implemented here yields statistically significant reductions in visibility estimation biases and uncertainties when compared to other methods. Our tests with CHARA/JouFLU data show that the improvements from using the median estimator are even greater for brighter stars, namely that the visibility bias is significantly smaller, as expected from the visibility model presented in Eq. \ref{model}. The tests presented above have been limited to mostly unresolved stars, but our results will likely be valid for resolved stars as long as the visibility is not extremely low and noisy (i.e. $|\rm{min}(V)|<\rm{Median(V)}$). Our motivation is to use these data reduction strategies in an upcoming paper describing the latest results of the exozodi survey using the CHARA/JouFLU instrument. But the methodology presented here applies to the estimation of interferometric visibilities in general, whether coming from single mode beam combiners or not, co-axial or multi-axial. It also generally applies to the interpretation of repeated measurements with poorly known noise characteristics. 

\section*{Acknowledgments:}
This research was supported by an appointment to the NASA Postdoctoral Program at the Jet Propulsion Laboratory administered by Universities Space Research Association under contract with NASA. PN and BM are grateful for support from the NASA Exoplanet Research Program element, though grant number NNN13D460T.
This work is based upon observations obtained with the Georgia State University Center for High Angular Resolution Astronomy Array at Mount Wilson Observatory. The CHARA Array is supported by the National Science Foundation under Grant No. AST-1211929. Institutional support has been provided from the GSU College of Arts and Sciences and the GSU Office of the Vice President for Research and Economic Development.
We also thank the anonymous referee for his/her valuable comments which improved the quality of this manuscript.

\section*{Appendix A: Other Visibility Estimators}

Here we provide the definitions of the visibility estimators used throughout this paper, which are described in \citet{theo_2005}. If we assume that the statistical distribution of the visibility is normal, then the following are unbiased estimators. $V_{\rm{norm}}$ is defined as 

\begin{equation}
  V_{\rm{norm}}= \left( \frac{3\langle |V|^2 \rangle^2 - \langle |V|^4 \rangle  }{2}\right)^{1/4},
\end{equation}

with the following uncertainty:

\begin{equation}
  \sigma_{\rm{norm}}^2 = \sqrt{\langle |V| \rangle^4 - \frac{1}{2}\left( \langle |V|^2 \rangle^2 - \langle |V|^4 \rangle \right)  } - \langle |V| \rangle ^2.
\end{equation}

The $V_{\rm{lognorm}}$ estimator is defined as:

\begin{equation}
  V_{\rm{lognorm}} = \exp\left(\lambda + \frac{1}{2}\sigma^2\right) ,  
\end{equation}

where 

\begin{equation}
  \lambda = \frac{1}{4}\rm{ln}\frac{\langle |V|^2 \rangle^4}{\langle |V|^4 \rangle},
\end{equation} 

\begin{equation}
  \sigma =  \frac{1}{4}\rm{ln}\frac{\langle |V|^4 \rangle}{\langle |V|^2 \rangle^2},
\end{equation}

and the corresponding uncertainty is 

\begin{equation}
  \sigma_{\rm{lognorm}}^2 = \exp{(2\lambda + 2 \sigma^2)}-\exp{(2\lambda+\sigma^2)}.
\end{equation}

\bibliography{drs_bibliography}

\begin{thebibliography}{}
\expandafter\ifx\csname natexlab\endcsname\relax\def\natexlab#1{#1}\fi

\bibitem[{{Absil} {et~al.}(2006){Absil}, {di Folco}, {M{\'e}rand}, {Augereau},
  {Coud{\'e} du Foresto}, {Aufdenberg}, {Kervella}, {Ridgway}, {Berger}, {ten
  Brummelaar}, {Sturmann}, {Sturmann}, {Turner}, \& {McAlister}}]{absil_2006}
{Absil}, O., {di Folco}, E., {M{\'e}rand}, A., {et~al.} 2006, A\&A, 452, 237

\bibitem[{{Absil} {et~al.}(2013){Absil}, {Defr{\`e}re}, {Coud{\'e} du Foresto},
  {Di Folco}, {M{\'e}rand}, {Augereau}, {Ertel}, {Hanot}, {Kervella},
  {Mollier}, {Scott}, {Che}, {Monnier}, {Thureau}, {Tuthill}, {ten Brummelaar},
  {McAlister}, {Sturmann}, {Sturmann}, \& {Turner}}]{absil_2013}
{Absil}, O., {Defr{\`e}re}, D., {Coud{\'e} du Foresto}, V., {et~al.} 2013, A\&
  A, 555, A104

\bibitem[{{Benson} {et~al.}(1995){Benson}, {Dyck}, \& {Howell}}]{benson}
{Benson}, J.~A., {Dyck}, H.~M., \& {Howell}, R.~R. 1995, AO, 34, 51

\bibitem[{{Boden}(2007)}]{boden_2007}
{Boden}, A.~F. 2007, NAR, 51, 617

\bibitem[{{Boyajian} {et~al.}(2012{\natexlab{a}}){Boyajian}, {McAlister}, {van
  Belle}, {Gies}, {ten Brummelaar}, {von Braun}, {Farrington}, {Goldfinger},
  {O'Brien}, {Parks}, {Richardson}, {Ridgway}, {Schaefer}, {Sturmann},
  {Sturmann}, {Touhami}, {Turner}, \& {White}}]{taby1}
{Boyajian}, T.~S., {McAlister}, H.~A., {van Belle}, G., {et~al.}
  2012{\natexlab{a}}, ApJ, 746, 101

\bibitem[{{Boyajian} {et~al.}(2012{\natexlab{b}}){Boyajian}, {von Braun}, {van
  Belle}, {McAlister}, {ten Brummelaar}, {Kane}, {Muirhead}, {Jones}, {White},
  {Schaefer}, {Ciardi}, {Henry}, {L{\'o}pez-Morales}, {Ridgway}, {Gies}, {Jao},
  {Rojas-Ayala}, {Parks}, {Sturmann}, {Sturmann}, {Turner}, {Farrington},
  {Goldfinger}, \& {Berger}}]{taby2}
{Boyajian}, T.~S., {von Braun}, K., {van Belle}, G., {et~al.}
  2012{\natexlab{b}}, ApJ, 757, 112

\bibitem[{{Choquet} {et~al.}(2014){Choquet}, {Menu}, {Perrin}, {Cassaing},
  {Lacour}, \& {Eisenhauer}}]{choquet_2014}
{Choquet}, {\'E}., {Menu}, J., {Perrin}, G., {et~al.} 2014, A\&A, 569, A2

\bibitem[{{Ciardi} {et~al.}(2001){Ciardi}, {van Belle}, {Akeson}, {Thompson},
  {Lada}, \& {Howell}}]{ciardi}
{Ciardi}, D.~R., {van Belle}, G.~T., {Akeson}, R.~L., {et~al.} 2001, ApJ, 559,
  1147

\bibitem[{{Coud{\'e} du Foresto} {et~al.}(2001){Coud{\'e} du Foresto},
  {Chagnon}, {Lacasse}, {Mennesson}, {Morel}, {Perrin}, {Ridgway}, \&
  {Traub}}]{fluor0}
{Coud{\'e} du Foresto}, V., {Chagnon}, G., {Lacasse}, M., {et~al.} 2001,
  Academie des Sciences Paris Comptes Rendus Serie Physique Astrophysique, 2,
  45

\bibitem[{{Coud{\'e} du Foresto} {et~al.}(1997){Coud{\'e} du Foresto},
  {Ridgway}, \& {Mariotti}}]{vincent_drs}
{Coud{\'e} du Foresto}, V., {Ridgway}, S., \& {Mariotti}, J.-M. 1997, A\&As,
  121, 379

\bibitem[{{Coud{\'e} du Foresto} {et~al.}(2003){Coud{\'e} du Foresto}, {Borde},
  {Merand}, {Baudouin}, {Remond}, {Perrin}, {Ridgway}, {ten Brummelaar}, \&
  {McAlister}}]{fluor2}
{Coud{\'e} du Foresto}, V., {Borde}, P.~J., {Merand}, A., {et~al.} 2003, in
  Society of Photo-Optical Instrumentation Engineers (SPIE) Conference Series,
  Vol. 4838, Interferometry for Optical Astronomy II, ed. W.~A. {Traub},
  280--285

\bibitem[{{Defr{\`e}re} {et~al.}(2012){Defr{\`e}re}, {Lebreton}, {Le Bouquin},
  {Lagrange}, {Absil}, {Augereau}, {Berger}, {di Folco}, {Ertel}, {Kluska},
  {Montagnier}, {Millan-Gabet}, {Traub}, \& {Zins}}]{defrere_2012}
{Defr{\`e}re}, D., {Lebreton}, J., {Le Bouquin}, J.-B., {et~al.} 2012, A\& A,
  546, L9

\bibitem[{{Di Folco} {et~al.}(2004){Di Folco}, {Th{\'e}venin}, {Kervella},
  {Domiciano de Souza}, {Coud{\'e} du Foresto}, {S{\'e}gransan}, \&
  {Morel}}]{di_folco_2004}
{Di Folco}, E., {Th{\'e}venin}, F., {Kervella}, P., {et~al.} 2004, \aap, 426,
  601

\bibitem[{{Ertel} {et~al.}(2014){Ertel}, {Absil}, {Defr{\`e}re}, {Le Bouquin},
  {Augereau}, {Marion}, {Blind}, {Bonsor}, {Bryden}, {Lebreton}, \&
  {Milli}}]{ertel_2014}
{Ertel}, S., {Absil}, O., {Defr{\`e}re}, D., {et~al.} 2014, A\&A, 570, A128

\bibitem[{{Hummel} {et~al.}(1998){Hummel}, {Mozurkewich}, {Armstrong},
  {Hajian}, {Elias}, \& {Hutter}}]{npoi_1998}
{Hummel}, C.~A., {Mozurkewich}, D., {Armstrong}, J.~T., {et~al.} 1998, AJ, 116,
  2536

\bibitem[{{Kervella} {et~al.}(2004){Kervella}, {S{\'e}gransan}, \& {Coud{\'e}
  du Foresto}}]{kervella}
{Kervella}, P., {S{\'e}gransan}, D., \& {Coud{\'e} du Foresto}, V. 2004, A\&A,
  425, 1161

\bibitem[{{Le Bouquin} {et~al.}(2011){Le Bouquin}, {Berger}, {Lazareff},
  {Zins}, {Haguenauer}, {Jocou}, {Kern}, {Millan-Gabet}, {Traub}, {Absil},
  {Augereau}, {Benisty}, {Blind}, {Bonfils}, {Bourget}, {Delboulbe},
  {Feautrier}, {Germain}, {Gitton}, {Gillier}, {Kiekebusch}, {Kluska},
  {Knudstrup}, {Labeye}, {Lizon}, {Monin}, {Magnard}, {Malbet}, {Maurel},
  {M{\'e}nard}, {Micallef}, {Michaud}, {Montagnier}, {Morel}, {Moulin},
  {Perraut}, {Popovic}, {Rabou}, {Rochat}, {Rojas}, {Roussel}, {Roux},
  {Stadler}, {Stefl}, {Tatulli}, \& {Ventura}}]{pionier}
{Le Bouquin}, J.-B., {Berger}, J.-P., {Lazareff}, B., {et~al.} 2011, A\&A, 535,
  A67

\bibitem[{{M{\'e}rand} {et~al.}(2005){M{\'e}rand}, {Bord{\'e}}, \& {Coud{\'e}
  du Foresto}}]{merand_2005}
{M{\'e}rand}, A., {Bord{\'e}}, P., \& {Coud{\'e} du Foresto}, V. 2005, A\&A,
  433, 1155

\bibitem[{{M{\'e}rand} {et~al.}(2006){M{\'e}rand}, {Kervella}, {Coud{\'e} du
  Foresto}, {Perrin}, {Ridgway}, {Aufdenberg}, {ten Brummelaar}, {McAlister},
  {Sturmann}, {Sturmann}, {Turner}, \& {Berger}}]{merand_2006}
{M{\'e}rand}, A., {Kervella}, P., {Coud{\'e} du Foresto}, V., {et~al.} 2006,
  A\&A, 453, 155

\bibitem[{Nordgren {et~al.}(2001)Nordgren, Sudol, \& Mozurkewich}]{npoi_2001}
Nordgren, T.~E., Sudol, J.~J., \& Mozurkewich, D. 2001, The Astronomical
  Journal, 122, 2707

\bibitem[{{Perrin}(1997)}]{perrin_1997}
{Perrin}, G. 1997, A\&A, 121, doi:10.1051/aas:1997323

\bibitem[{{Perrin}(2003)}]{perrin_2003}
---. 2003, A\&A, 398, 385

\bibitem[{{Perrin} {et~al.}(1998){Perrin}, {Coud{\'e} du Foresto}, {Ridgway},
  {Mariotti}, {Traub}, {Carleton}, \& {Lacasse}}]{perrin_temp}
{Perrin}, G., {Coud{\'e} du Foresto}, V., {Ridgway}, S.~T., {et~al.} 1998,
  A\&A, 331, 619

\bibitem[{{Richichi} {et~al.}(2005){Richichi}, {Percheron}, \&
  {Khristoforova}}]{richichi}
{Richichi}, A., {Percheron}, I., \& {Khristoforova}, M. 2005, A\& A, 431, 773

\bibitem[{{Roddier} \& {Lena}(1984)}]{roddier_1984}
{Roddier}, F., \& {Lena}, P. 1984, Journal of Optics, 15, 171

\bibitem[{{Scott} {et~al.}(2013){Scott}, {Millan-Gabet}, {Lhom{\'e}}, {Ten
  Brummelaar}, {Coud{\'e} Du Foresto}, {Sturmann}, \& {Sturmann}}]{nic}
{Scott}, N.~J., {Millan-Gabet}, R., {Lhom{\'e}}, E., {et~al.} 2013, Journal of
  Astronomical Instrumentation, 2, 40005

\bibitem[{{Shao} {et~al.}(1988){Shao}, {Colavita}, {Hines}, {Staelin},
  {Hutter}, {Johnston}, {Mozurkewich}, {Simon}, {Hershey}, {Hughes}, \&
  {Kaplan}}]{shao_1988}
{Shao}, M., {Colavita}, M.~M., {Hines}, B.~E., {et~al.} 1988, A\&A, 193, 357

\bibitem[{{Tango} \& {Twiss}(1980)}]{tango_and_twiss}
{Tango}, W.~J., \& {Twiss}, R.~Q. 1980, in Progress in optics. Volume 17.
  (A81-13109 03-74) Amsterdam, North-Holland Publishing Co., 1980, p. 239-277.
  Research supported by the Australian Research Grants Committee., ed.
  E.~{Wolf}, Vol.~17, 239--277

\bibitem[{{ten Brummelaar} {et~al.}(2005{\natexlab{a}}){ten Brummelaar},
  {McAlister}, {Ridgway}, {Bagnuolo}, {Turner}, {Sturmann}, {Sturmann},
  {Berger}, {Ogden}, {Cadman}, {Hartkopf}, {Hopper}, \& {Shure}}]{chara}
{ten Brummelaar}, T.~A., {McAlister}, H.~A., {Ridgway}, S.~T., {et~al.}
  2005{\natexlab{a}}, ApJ, 628, 453

\bibitem[{{ten Brummelaar} {et~al.}(2005{\natexlab{b}}){ten Brummelaar},
  {McAlister}, {Ridgway}, {Bagnuolo}, {Turner}, {Sturmann}, {Sturmann},
  {Berger}, {Ogden}, {Cadman}, {Hartkopf}, {Hopper}, \& {Shure}}]{theo_2005}
---. 2005{\natexlab{b}}, ApJ, 628, 453

\bibitem[{{White} {et~al.}(2013){White}, {Huber}, {Maestro}, {Bedding},
  {Ireland}, {Baron}, {Boyajian}, {Che}, {Monnier}, {Pope}, {Roettenbacher},
  {Stello}, {Tuthill}, {Farrington}, {Goldfinger}, {McAlister}, {Schaefer},
  {Sturmann}, {Sturmann}, {ten Brummelaar}, \& {Turner}}]{white_2013}
{White}, T.~R., {Huber}, D., {Maestro}, V., {et~al.} 2013, MNRAS, 433, 1262

\end{thebibliography}
\end{document}